# When wall slip wins over shear flow: A temperature-dependent Eyring slip law and a thermal multiscale model for diamond-like carbon lubricated by a polyalphaolefin oil

*Stefan Peeters[*a,b], Edder J. García[a,b], Franziska Stief[a,c], Thomas Reichenbach[a], Kerstin Falk[a], Gianpietro Moras[a], Michael Moseler[**a,b,c]*

[a]Fraunhofer IWM, MikroTribologie Centrum µTC, Wöhlerstraße 11, 79108 Freiburg, Germany

[b]Freiburg Materials Research Center, University of Freiburg, Stefan-Meier-Straße 21, 79104 Freiburg, Germany

[c]Institute of Physics, University of Freiburg, Herrmann-Herder-Straße 3, 79104 Freiburg, Germany

# Abstract

The quantitative description of lubricant flow in nanoscale channels is complicated by various finite-size effects that are not taken into account in conventional thermo-elasto-hydrodynamic lubrication (TEHL) models. One of these effects is wall slip, a phenomenon that has been extensively studied both theoretically and experimentally. The relationship between wall slip and thermal effects is intricate, and some authors debate whether the friction reduction observed in their experiments in the TEHL regime can be explained by either slip or viscosity reduction in heated lubricants. To disentangle these mechanisms, a comprehensive molecular dynamics study of the relationship between temperature and slip in the shear flow of a 4 cSt polyalphaolefin (PAO4) base oil in a nanoscale diamond-like carbon (DLC) channel is performed here. An Eyring law describes the relationship between slip velocity and shear stress at the solid-liquid interface for a given temperature and pressure. The same simulation campaign provides a pressure-dependent law for the interface thermal resistance (ITR) between DLC and PAO4. These constitutive laws are employed in a continuum model for lubricated parallel channels. By taking heat conduction into account and combining the slip and ITR laws with a temperature- and pressure-dependent Eyring viscosity law, questions about the competition of slip and thermal thinning of the lubricant can be answered. For DLC film thicknesses compatible with tribological experiments and applications, this model shows that slip is only relevant for very thin lubricant films that are typical of boundary lubrication, suggesting the dominance of thermal thinning in the TEHL regime.

# Introduction

Wall slip in a lubricated channel occurs whenever the speed of a solid material and the average velocity of the lubricant molecules in direct contact with the solid differ. Slip affects the velocity profile, hence the shear rate, of the lubricant in the channel and can reduce friction in the TEHL regime[1,2]. This phenomenon is particularly relevant for nanosized channels[3], which are becoming increasingly common due to the technological progress towards miniaturization and lower lubricant viscosities, favoring the transition to the mixed and boundary lubrication regimes. In these systems, the properties of the lubricant can deviate from those of the bulk fluid due to the emergence of finite-size effects, for instance changes in viscosity and density layering[4].

In a recent work, we have demonstrated the importance of slip to describe lubricant flow in nanosized tribological contacts using a continuum model[5]. The Reynolds lubrication equation (RLE), the most common method to describe fluid flow in slender channels, typically relies on the no-slip boundary condition. We have shown that this assumption is not valid for the flow of hexadecane on Au(111) as well as PAO4 on H-terminated DLC surfaces at high pressures or for nanometer-thin lubricant films. In our work, we have used a constitutive slip law parametrized with molecular dynamics (MD) simulations and combined it with constitutive laws for the density and viscosity to inform an extended RLE continuum model. For the material combinations Au/hexdecane and DLC/PAO4 considered in Ref. 5, this has led to a multiscale description of boundary lubrication consistent with our large-scale atomistic benchmark simulations.

For the sake of simplicity, this initial model was based on the isothermal RLE, neglecting temperature effects that influence slip. However, heat conduction and related thermal variations in lubricants and solids are relevant for many applications including machining[6] and grinding[7].

Moreover, they are the subject of active research owing to the increasing number of electric vehicles (EV) on the market, since lubricants for EVs are selected and designed not only for their ability to reduce friction and wear, but also for their thermal properties[8].

Energy losses in tribological contacts crucially depend on the dominant mode of sliding velocity accommodation, which can occur either by shearing the lubricant or by slip of the lubricant on solid surfaces. Since slip and viscosity vary with temperature, viscous heating could tip the balance between these two energy dissipation modes. Because of this subtle competition, different ideas have been proposed in the literature to explain low friction in the TEHL regime. For instance, Björling et al.[9,10] performed tribological tests using DLC-coated steel balls and discs and compared the results with their uncoated counterparts. They observed a lower friction coefficient and a higher temperature in the PAO4 lubricant compared to the case of the uncoated material, and suggested that the low thermal conductivity of DLC is responsible for low friction, since viscous heating in the lubricant reduces its viscosity[9,10]. Conversely, Kalin et al. performed similar experiments[2,11,12] and identified slip as a key factor to determine low friction of DLC, excluding thermal effects[13].

To answer the question whether the friction reduction observed experimentally on DLC can be mainly attributed to wall slip, thermal viscosity reduction, or a combination of both, it is essential to quantify the temperature-dependence of constitutive laws entering the RLE. In particular, an accurate temperature-dependent slip law is needed. Most likely, the interfacial thermal resistance (ITR, a quantity that indicates the resistance of heat transfer across an interface) is also necessary to develop an extended TEHL model that can disentangle the impacts of temperature-induced changes in slipperiness and viscosity on friction. Therefore, it is essential to extend our previous constitutive law for slip to include temperature dependence, alongside the pressure dependence, of the characteristic parameters of slip. For this purpose, we performed a systematic MD study of 1-decene trimers ($C_{10}$-trimer or $C_{30}H_{62}$, the main component of PAO4), flowing between two hydrogen-terminated DLC surfaces and considered different thermalization schemes and temperatures. In the following, we report the construction of our temperature- and pressure-dependent slip law, as well as an empirical pressure-dependent law for the ITR between the lubricant and the DLC surface. We have two objectives in mind: (i) to quantify the effect of temperature differences and inhomogeneities on slip and (ii) to estimate for which experimental conditions slip dominates over shear flow in the lubricant. To achieve the second objective, the laws for slip and ITR, as well as an additional viscosity law, are used in a simple thermal continuum model (TCM) for a parallel channel. The TCM allows for an assessment of the relevance of wall slip for PAO4 and DLC film heights beyond those considered in the MD campaign.

## Materials and Methods

**Molecular dynamics**

All the MD simulations were performed using LAMMPS[14]. A Velocity-Verlet algorithm with a time step of 0.5 fs was used to integrate the equations of motion. Parallel channels of amorphous carbon were lubricated by $C_{30}H_{62}$, in analogy to our previous work[5]. The systems contained 88 lubricant molecules and two amorphous carbon (a-C) slabs representing DLC with a density $\rho_{\text{DLC}} = 2.5$ g/cm$^3$. The total number of atoms in the system was 24754. A representative geometry of the systems, visualized with OVITO[15], is shown in Figure 1a. The parallel channels were enclosed in orthorhombic cells with dimensions $L_x$ = 8 nm, $L_y$ = 4 nm, and $L_z \approx 6.2$ nm. Periodic boundary conditions were adopted along the $x$ and $y$ directions. The lubricant film thickness or gap height $h$, which represents the distance between the surfaces of the two amorphous carbon walls, was defined as the distance along $z$ between the two lowest minima in the total density profile in the system and was approximately 2 nm for all the parallel channels.

Molecular interactions were described by the L-OPLS potential[16], while bulk and surface atoms of the DLC slabs were described using an OPLS potential recently developed in our group to represent the geometry and the elastic constants of a-C structures, as well as their H- and OH-passivated surfaces[17]. Our OPLS potential is interfaced to LAMMPS via matscipy[18] and ASE[19]. Interactions between slabs and fluid were determined by geometry-mean combination rules, following the OPLS convention[20]. Long-range electrostatic interactions were described using the particle-particle particle-mesh (PPPM) algorithm[21] implemented in LAMMPS, with an accuracy of $10^{-5}$ eV/Å on the forces.

The initial geometry of the a-C slabs was obtained after a melt-quench protocol already adopted in our previous works[5,17,22,23]. The surfaces of the slabs in contact with the fluid were passivated by saturating surface C atoms with H atoms. The atoms belonging to the top- and bottommost regions of the parallel channel were kept rigid and fixed, respectively. The velocity of the fixed atoms was set to zero, while a constant velocity $u_x$ was applied to the rigid atoms. The thickness of these constrained regions was 0.5 nm. The lubricated interfaces were equilibrated for 0.5 ns, and the pressure in the system was kept constant via the pressure-coupling algorithm by Pastewka, Moser and Moseler[24]. The simulations were carried out for 17 different values of the sliding velocity $u_x$ (1, 2.5, 5, 7.5, 10, 15, 20, 30, 40, 50, 60, 70, 80, 90, 100, 150, 200 m/s) and 5 different values of pressure $P$ (0.2, 0.5, 1.0, 1.5, 2.0 GPa). In the following, only the results up to $u_x =$ 100 m/s are shown, while higher sliding velocities for specific systems are presented only in the Supporting Information (SI). During the second half of the equilibration, the average height of the system was calculated, then the position of the top wall was gradually adjusted to match the average height. For this purpose, a constant velocity along the vertical direction was imposed on the rigid atoms of the top wall for 60 ps. Finally, sliding simulations were run at constant height for 5 ns to calculate the slip properties. During the dynamic simulations, a Langevin thermostat was coupled to the *y*-velocities of all unconstrained atoms in the system, i.e., in the solids and in the lubricant. The target temperature of the thermostat was set to 250, 300, 350, 400 or 450 K (to cover a range of temperatures typically observed in EV drivetrains and combustion engines) and 0.1 ps was chosen as the time constant of the thermostat. We refer to this thermostatting scheme, that ensures fully isothermal conditions, as TWTF (thermostat walls, thermostat fluid), to follow the convention adopted in similar computational works[25,26]. An alternative thermostatting scheme called TW (thermostat walls) was also considered for the sliding simulations at 350 K (as well as 250 and 450 K, only in the SI). In the TW setup, the Langevin thermostat was applied in two 0.6 nm-thick regions adjacent to the fixed and rigid regions in the DLC slabs. The temperature of the remaining atoms was not controlled in this setup. The two thermostatting approaches are schematically shown in Figure 1b-c. High sliding velocities in the non-isothermal setup inevitably produced higher temperatures in the lubricant than in the a-C due to viscous heating[25,27]. To better evaluate the role played on slip by a warm lubricant on cool surfaces, we performed a third set of simulations using the TWTF scheme in which the temperature of the DLC and of the lubricant were set to 350 K and 425 K, respectively. The artificial discontinuity in temperature forced by this non-isothermal TWTF (NI-TWTF) scheme might be unrealistic, yet it still is a useful model to distinguish the contributions of the temperature of the solid slabs and of the lubricant to slip. The NI-TWTF simulations were compared with an additional TWTF system at $T = 425$ K and $P = 1$ GPa. In total, 578 combinations of sliding velocity, temperature, thermostatting scheme and pressure were considered in this work, resulting in as many independent MD simulations. A summary of the calculated properties for all simulation systems is provided as SI.

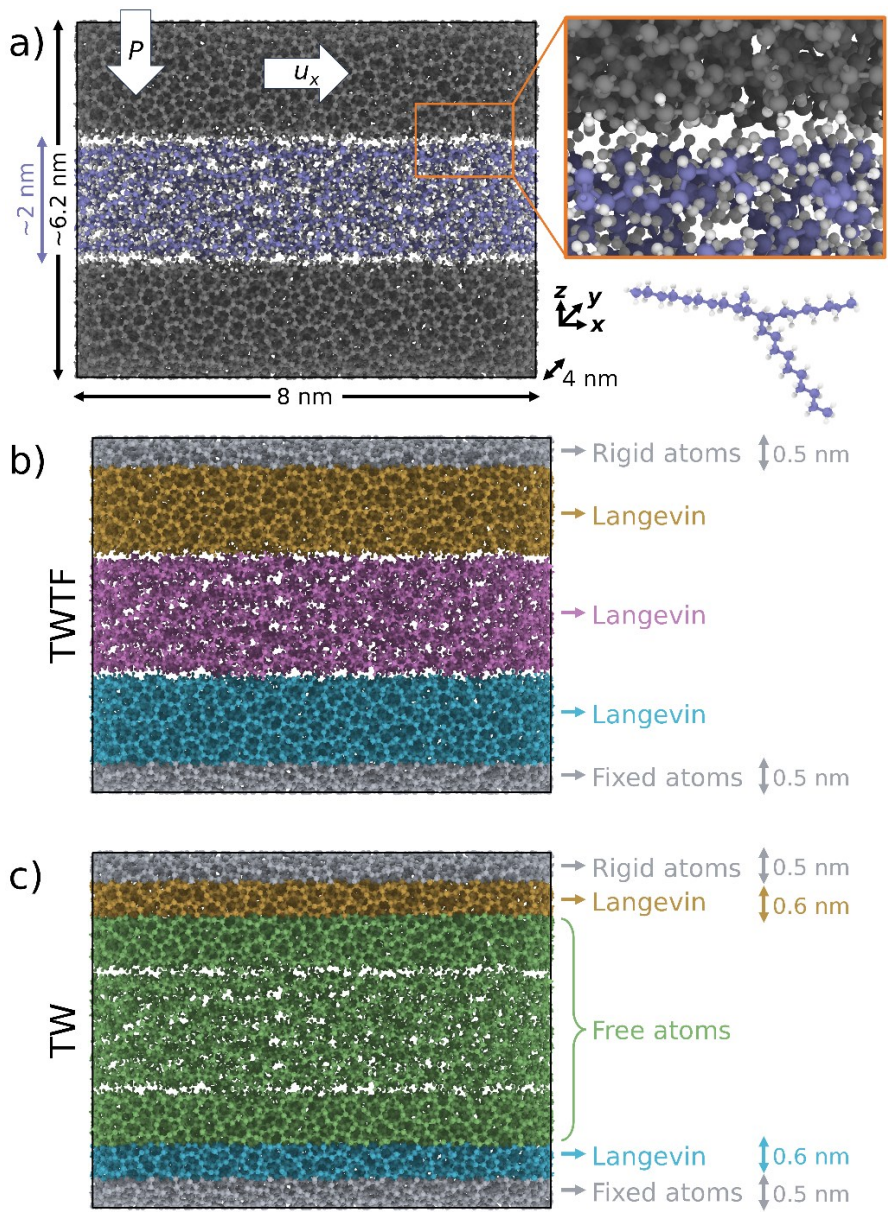

Figure 1 – (a) Lateral view of the simulation setup. Note that pressure $P$ is only applied during equilibration, since the slip properties are calculated at constant height. The inset shows a close-up of the lubricant/slab interface. A molecule of $C_{30}H_{62}$ is shown below the inset. Panels b and c display the setup for the TWTF and TW simulations, respectively.

**Calculation of slip, viscosity, and thermal properties**

Density, velocity, and temperature profiles along $z$ were obtained during the sliding simulations. For each of the 5 ns production runs, the velocity profiles averaged over intervals of 1 ns were fitted with a Couette profile using the least-squares method, providing a set of 5 shear rates $\dot{\gamma}$ for a given sliding velocity $u_x$. The data points in two 0.5 nm-regions in the lubricant in contact with the surfaces were removed from the fits, to consider only the linear part of the velocity profile. The extrapolation of the Couette profiles identified 10 intersections with the slabs (5 each for the top and bottom surfaces, where the density profiles reached the lowest minima between the a-C slabs), providing the slip velocities $v_{x1}^s$ and $v_{x2}^s$. The overall slip velocity was taken as the average of the absolute values of the slip velocities at both surfaces, $v_s = \frac{|v_{x1}^s| + |v_{x2}^s|}{2}$. The shear stress $\tau$ at the surfaces was obtained by calculating the interaction force between the PAO4 and the DLC in the sliding direction ($x$) divided by the surface area of the slabs. The dependence of $v_s$ on $\tau$ was fitted using the Eyring constitutive law[28], which is based on the molecular kinetic theory and describes slip as an activated process[29]:

$$v_s(\tau, P, T) = v_0(P, T) \sinh\left(\frac{\tau}{\tau_0(P,T)}\right), \qquad (1)$$

where the fitting parameters $v_0(P, T)$ and $\tau_0(P, T)$ are the characteristic velocity and shear stress, respectively. Initially, the slip velocities in this work were fitted according to Eq. (*1*) for each individual $(P, T)$ combination followed by a subsequent fit of $v_0(P, T)$ and $\tau_0(P, T)$ by suitable semi-empirical relationships. However, as explained in the Results and Discussion section, more sophisticated fitting strategies must be used to reduce the uncertainty in the fitted parameters. Such strategies will be briefly introduced whenever adopted and are explained in more detail in the SI (Section S1). The orthogonal distance regression (ODR) algorithm implemented in SciPy[30] was adopted for all fitting procedures of Eq. (*1*).

In the TW simulations, we also calculated the interfacial thermal resistance $R_{th}$, which indicates the resistance of heat transfer across an interface and can be obtained in the presence of a temperature discontinuity between two phases. The ITR was calculated as:

$$R_{th} = \frac{\Delta T}{J_Q}, \qquad (2)$$

where $\Delta T$ is the temperature jump between the surface and the lubricant and $J_Q$ is the heat flux across the interface. The temperature jump $\Delta T$ was calculated as the difference between the temperatures of the surfaces $T_s$ and the lubricant $T_l$ at the interface. These quantities were obtained by three separate least-squares fits of the temperature profile, namely a linear fit for each slab and a parabolic fit for the lubricant. In each fit, the data points in 0.5 nm regions at the lubricant-slab interface were discarded, as well as 0.55 nm-thick regions to exclude the frozen and rigid atoms from the slabs. The temperature of the lubricant at the interface $T_l$ was taken as the average of the values obtained from intersecting the parabola with the top and the bottom interfaces, and the same was done for the fitted straight lines to obtain $T_s$ (as shown in Figure 2). The heat flux $J_Q$ was calculated by averaging the total heat transferred by the two Langevin thermostats and normalizing this quantity by the area of the base of the simulation cell (equal to 24 nm²) and the simulation time. As in the case of the slip properties, the ITR was calculated for each nanosecond of the simulation to produce a set of 5 values, which were then averaged.

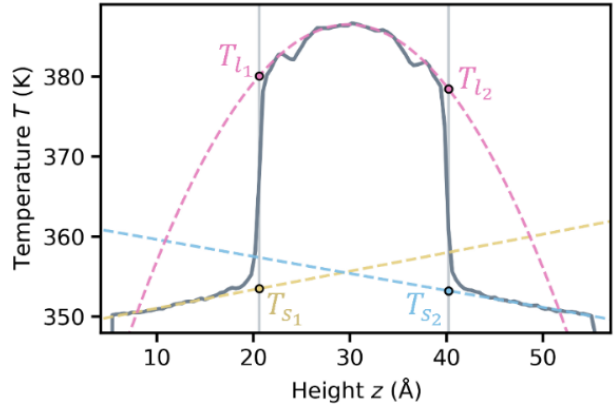

Figure 2 – Example of a temperature profile (TW system at $P = 2$ GPa and $u_x = 100$ m/s) with the linear and parabolic fits to determine $\Delta T = (T_{l_1} + T_{l_2} - T_{s_1} - T_{s_2})/2$.

## Results and Discussion

**Determining the temperature dependence of the slip velocity – shear stress relationship**

In our previous work, an Eyring slip law was parametrized for a single temperature $T = 400$ K using isothermal MD[5]. In the following, an extensive isothermal simulation campaign is performed for a set of $T$ covering a typical temperature range of tribologically loaded DLC coatings. The TWTF scheme is used here to study the flow of $C_{30}H_{62}$ between the two parallel surfaces of H-terminated DLC at $T = 250, 300, 350, 400$ and $450$ K and $P = 0.2, 0.5, 1.0, 1.5$ and $2.0$ GPa. Figure 3 shows that the slip velocity $v_s$ has a strong dependence on pressure $P$ and shear stress $\tau$, while the variation of $v_s$ with temperature $T$ is comparatively small. The fitted Eyring laws (with individual $\tau_0$ and $v_0$ for each $(P, T)$ combination) represents the MD data points well, despite some statistical error on the calculated $v_s$ and $\tau$ (the standard deviations of $v_s$ and $\tau$ are visible in Figure S6 in the SI, Section S2).

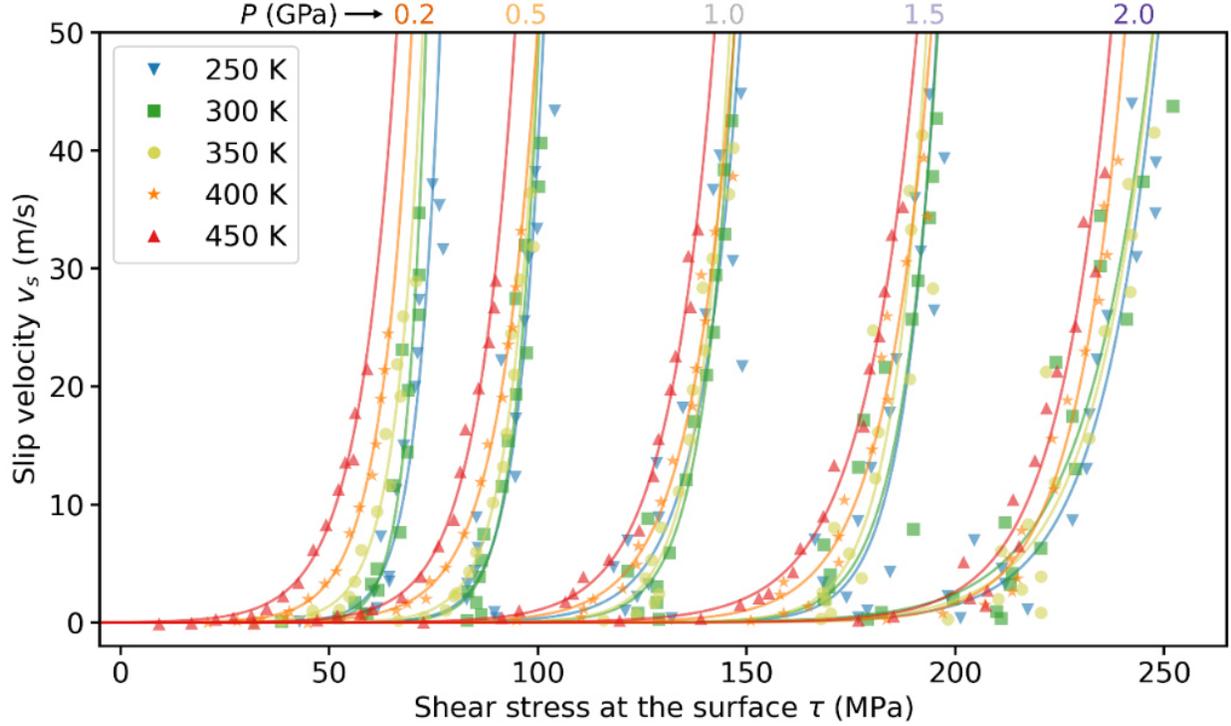

Figure 3 – Slip velocities as a function of the shear stress at the DLC surface for the isothermal simulations (TWTF) at 250, 300, 350, 400 and 450 K (data points) and the corresponding slip laws according to Eq. (1) (solid lines). The five groups of curves identify different pressures from 0.2 to 2 GPa as indicated at the top.

**Parametrization of a temperature- and pressure-dependent slip law**

It is worth to briefly discuss the reason why, for given shear stress $\tau$ and pressure $P$, the slip velocity increase with an increase in temperature, as shown in Figure 3. The excellent agreement between the MD data and the Eyring law[28] (Eq. (1)) indicates that slip is an activated process, as proposed by several authors[31–33]. The characteristic parameters of the slip law can be written as:

$$\tau_0(T) = \frac{2k_B T}{\lambda S} \qquad (3)$$

and

$$v_0(T) = \frac{2\lambda k_B T F_{\text{act}}}{h_P F_{\text{in}}} e^{-\frac{E_0}{k_B T}}, \quad (4)$$

where $\lambda$ is the length of a slip event (a geometric parameter related to the energy corrugation of the surface), $S$ is the corresponding activation area along the sliding direction, $F_{\text{act}}$ and $F_{\text{in}}$ are the partition functions of the activated (slipping) and initial (sticking) states, respectively, $E_0$ is the activation energy required for an individual slip event of a lubricant molecule, and $k_B$ and $h_P$ are the Boltzmann and the Planck constants. The effect of temperature on the characteristic shear stress $\tau_0(T)$ is straightforward, as an increase in $T$ leads to a decrease of the $\tau/\tau_0$ term in the hyperbolic sine of Eq. (1), moving the onset of the slip regime linearly to higher shear stresses. The characteristic velocity $v_0(T)$ also increases with temperature, although nonlinearly, and this counteracts the contribution of $\tau_0(T)$. In our systems, the slip curves shift to lower shear stresses with increasing temperature, indicating that $v_0(T)$ plays a more important role than $\tau_0(T)$ on the slip law.

Since we derived a set of isothermal slip laws at different temperatures (and pressures), Eqs. (3) and (4) can be used to fit these results and obtain the temperature dependence of slip. This fit strategy turned out to be inadequate to provide values of $\tau_0$ and $v_0$ in the Eyring law (as shown in Figure S4). This issue does not arise because of the scattering in the MD data, but because both $\tau_0$ and $v_0$ concurrently change the shape and the position of the slip curve. In this case, a constrained fit not only reduces the uncertainty on the characteristic parameters but also provides a better quality of the fit with the Eyring law. The details of the constrained fit adopted here are presented in the SI. In short, we first derive $\tau_0(T)$ for each pressure (Figure 4a). According to Eq. (3), $\tau_0(T)$ is determined by a single parameter, i.e., the slope $c_1 = \frac{2k_B}{\lambda S}$ in the linear relation $\tau_0(T) = c_1 T$. The pressure dependence of this parameter can also be fitted linearly (Figure 4b):

$$c_1(P) = A_{c_1}P + B_{c_1}. \tag{5}$$

Using a set of values of $c_1$ that lie on $c_1(P)$, we can refit the slip laws (Eq. (1)) to improve the estimation of $v_0$. In a second iteration of the constrained fit, Eq. (4) is used and $v_0$ is fitted by $v_0(T) = c_2 \exp(-c_3/T)$ (see Figure 4a). This provides the coefficient $c_2 = \frac{2\lambda k_B F_{act}}{hF_{in}}$ and the activation temperature $c_3 = \frac{E_0}{k_B}$. The dependence on pressure of both parameters can be fitted with an exponential function and a parabola, respectively (Figure 4c and d):

$$c_2(P) = A_{c_2} e^{B_{c_2} P}, \tag{6}$$

$$c_3(P) = A_{c_3} P^2 + B_{c_3} P + C_{c_3}. \tag{7}$$

Constraining $c_2$ and $c_3$ to lie on the corresponding pressure-dependent curves allows to refine $\tau_0$, and therefore $c_1(P)$, in a third iteration. A final fit of the Eyring laws with such a refined $c_1(P)$ provides a set of seven parameters reported in Table 1 that can be used to calculate slip for arbitrary combinations of temperature ($300 \text{ K} \leq T \leq 450 \text{ K}$) and pressure ($0 < P \leq 2 \text{ GPa}$). This is effectively a temperature- and pressure-dependent constitutive slip law for the $C_{10}$-trimer on H-terminated DLC, which is useful to determine slip for this lubricant/surface combination without the need to perform additional MD simulations.

Table 1 – Parameters obtained from the final fit of Eqs. (5)-(7).

| Parameter | Value |
|---|---|
| $A_{c_1}$ | $9.4928 \cdot 10^{-6} \frac{1}{\text{K}}$ |
| $B_{c_1}$ | $1.5783 \cdot 10^{-2} \frac{\text{MPa}}{\text{K}}$ |
| $A_{c_2}$ | $(17.417 \pm 3.931) \frac{\text{m}}{\text{s} \cdot \text{K}}$ |
| $B_{c_2}$ | $(-2.1102 \pm 0.2227) \cdot 10^{-3} \frac{1}{\text{MPa}}$ |
| $A_{c_3}$ | $(-4.4178 \pm 0.1501) \cdot 10^{-4} \frac{\text{K}}{\text{MPa}^2}$ |
| $B_{c_3}$ | $(1.8509 \pm 0.0419) \frac{\text{K}}{\text{MPa}}$ |
| $C_{c_3}$ | $(5.1766 \pm 0.0269) \cdot 10^3 \text{ K}$ |

The agreement of the predicted slip laws with the MD results is good (Figure 4e) given the wide range of temperatures and pressures considered. For few low-temperature and/or high-pressure systems the model predicts lower slip than the one in the MD. As we will show in the following, this slip model can be implemented in continuum models to update the slip properties of the lubricant fluid when the temperature of the system is expected to change. In this context, we shall briefly comment on the choice of the functional forms in Eqs. (5)-(7). Although Eqs. (5) and (6) are empirical, they are monotonic and represent well the pressure dependence of $\tau_0$ and $v_0$. Future work should nevertheless focus on providing an equally accurate, physically motivated dependence. Conversely, Eq. (7) can be related to the pressure dependence of the Gibbs free energy of activation at constant temperature, $\Delta G^\ddagger(P) = \Delta G_0^\ddagger + \Delta V_0^\ddagger P + \frac{\Delta \kappa^\ddagger}{2} P^2$, where $\Delta G_0^\ddagger$ and $\Delta V_0^\ddagger$ are respectively the activation energy and volume at $P = 0$ and $\Delta \kappa^\ddagger$ is the activation compressibility[34].

In the SI (Table S3 and Figure S5), we also provide an alternative parametrization in which $c_3(P)$ has a logarithmic dependence on $P$. This alternative formulation yields a slightly worse fit quality compared to the one described above (i.e., by using Eq. (7)) but ensures a monotonic behavior consistent with $c_1(P)$ and $c_2(P)$ without significant changes to the predicted slip laws.

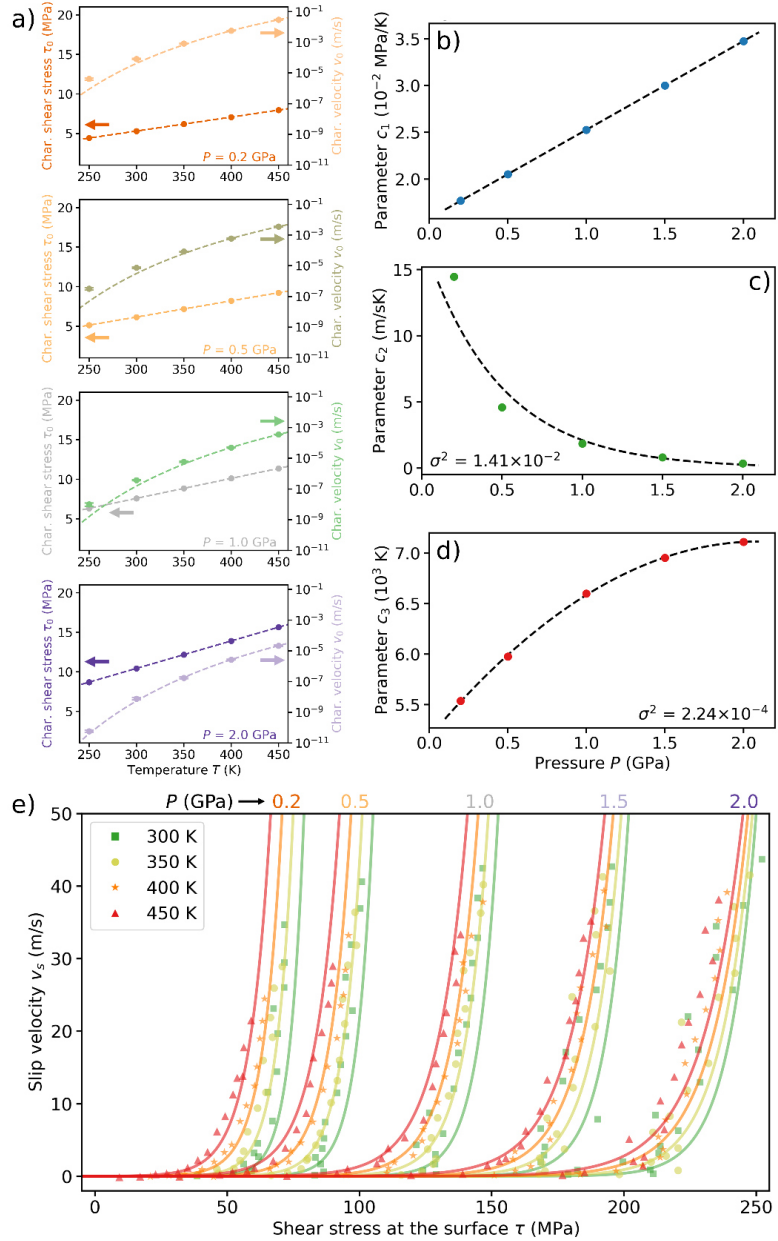

Figure 4 – (a) $\tau_0(T,P)$ and $v_0(T,P)$ fitted by Eqs. (3) and (4) using the fitting procedure described in the text. The barely visible error bars on $v_0$ represent the uncertainty of the estimation of these parameters. (b)-(d) show the respective fit of $c_1(P)$, $c_2(P)$ and $c_3(P)$ with Eqs. (5)-(7). $\sigma^2$ represents the residual variance of the corresponding ODR fits. (e) MD results for $T \geq 300$ K and the slip law at the corresponding temperatures and pressures predicted by the model.

**Pressure-dependent law for the interfacial thermal resistance**

Another crucial quantity that describes the thermal coupling between lubricant and solid is the interfacial thermal resistance, known as Kapitza resistance for crystalline surfaces[35]. High values of the ITR indicate a reduced momentum transfer between the surface and the lubricant molecules[36], an aspect that is also highly relevant for slip and thermal thinning of the lubricant. In continuity with the simulations described so far, we investigated changes in the ITR using the same sliding setup that considers two thermostatting schemes: the TWTF scheme, in which a Langevin thermostat controls the temperature of all movable atoms in the a-C slabs and the lubricant, and the TW scheme, in which the Langevin thermostat is applied only to thin regions in the DLC slabs adjacent to the zones with rigid C atoms. In the TW setup, a temperature difference originates between the thermostatted regions and the lubricant due to viscous heating, the magnitude of which depends on pressure $P$ and sliding velocity $u_x$[25,37] (up to 36 K in our simulations). Figure 5 shows the slip laws of the TWTF and TW systems at different pressures. A comparison between the filled (TWTF) and the empty (TW) symbols suggest that the choice of the thermostatting scheme has some impact on the slip properties for this combination of lubricant and substrate, as the fitted Eyring constitutive laws of the TW system are shifted consistently to smaller shear stress compared to the TWTF case.

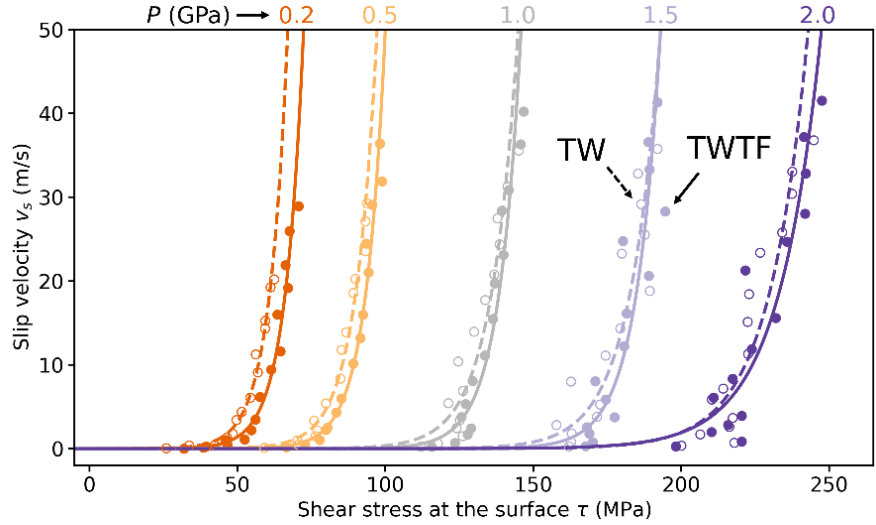

Figure 5 – Slip velocities as a function of the shear stress for the TWTF (filled circles and solid lines) and TW (empty circles and dashed lines) simulations at $T = 350$ K and the corresponding slip laws fitted to Eq. (1) without constraints.

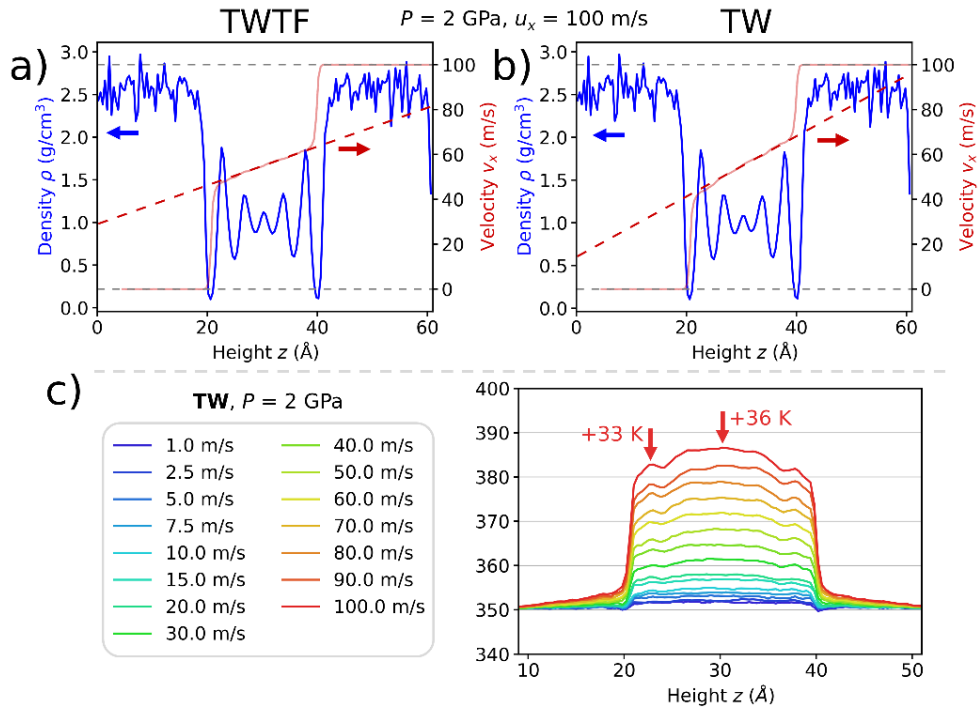

Figure 6 – Density and velocity profiles for the TWTF (a) and TW (b) systems at $P = 2$ GPa and $u_x = 100$ m/s. (c) Temperature profiles of the TW systems at $P = 2$ GPa for different sliding velocities $u_x$.

The density profiles do not differ significantly in the two approaches even for the system with $P = 2$ GPa and $u_x = 100$ m/s (Figure 6a and b). The velocity profile of the TWTF systems, however, is characterized by a smaller shear rate compared to the corresponding TW systems, leading to larger slip. The temperature profile shows a moderate temperature increase of 33 K in the lubricant in contact with the surface (36 K at the center of the lubricant layer) compared to the target temperature of 350 K, as well as a slight temperature gradient in the DLC slabs (Figure 6c). The discontinuity in the temperature profile allows the calculation of the ITR, which is shown for all the TW systems in Figure 7. Interestingly, this quantity is effectively constant for $u_x \geq 40$ m/s (and for $T = 250\text{-}450$ K, as shown in Figure S7). This suggests that the thermal coupling between lubricant and surface does not change significantly above a certain sliding velocity. The ITR at low sliding speed is higher than the one at high speed, but so is the uncertainty and therefore, the significance of those datapoints. The inset shows that the ITR is deeply connected to the pressure (and, therefore, the shear stress) experienced by the fluid in contact with the wall, since the average values of the ITR for each pressure collapse on a single curve that can be fitted remarkably well with an exponential law $R_{\text{th}}(P) = A_P e^{B_P P} + C_P$. This result offers an intuitive interpretation for the behavior of the ITR of a sheared fluid, i.e., the thermal conductivity increases with the pressure, and therefore with the interfacial stress, a characteristic measure of the momentum transfer between the lubricant and solid walls.

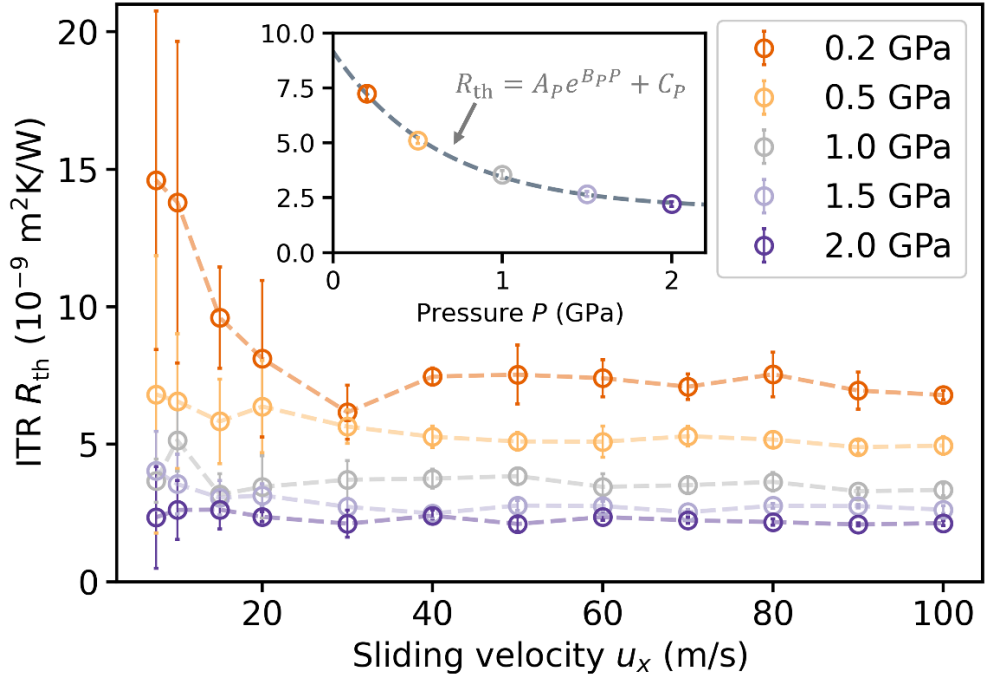

Figure 7 – Interfacial thermal resistance ITR of the TW systems as a function of sliding velocity and pressure. The data points between $u_x$ = 1 and 2.5 m/s were discarded in the main plot due to the large uncertainty of the resulting ITR. The error bars represent the standard deviation of 5 individual evaluations for each system. The dashed lines in the main plot are a guide to the eye, while the grey dashed line in the inset represents the fitted exponential law. The error bars in the inset mark the standard deviation of the ITR averaged in the range 40-100 m/s. The parameters resulting from the ODR fits are $A_P = 7.20 \cdot 10^{-9}$ m²K/W, $B_P = -1.58$ GPa$^{-1}$ and $C_P = 1.96 \cdot 10^{-9}$ m²K/W.

**Building the TCM: Choosing the appropriate temperature in the slip law**

To estimate for which experimental conditions slip dominates over shear flow in the lubricant, we developed a simple non-isothermal continuum model, the TCM. This model should be able to reproduce the shear stresses, slip properties, temperatures in the solid and lubricant from the MD. Since we observed a pronounced temperature discontinuity at the solid-lubricant interface due to the non-zero ITR in our simulations, it is not trivial to answer the question whether the lubricant temperature, the solid temperature or an average of both determines the magnitude of slip. To address this issue, and understand which temperature should be chosen for the TCM, we performed an additional set of non-isothermal (NI-TWTF) simulations where the lubricant was thermostatted to 425 K while the slabs were kept at 350 K, as well as a set of simulations with a TWTF system at 425 K. Slip under the NI-TWTF conditions was compared to the one in the TWTF simulations at 350 and 425 K. As shown in Figure 8, the slip laws of the TWTF simulations at 350 K are shifted to higher shear stresses than in the NI-TWTF systems, while the TWTF system at 425 K shows only slightly more slip than the NI-TWTF simulations, despite the 75 K difference in the wall temperature. These results indicate that, unlike the lubricant temperature, the temperature of the DLC surfaces has little effect on slip in these systems.

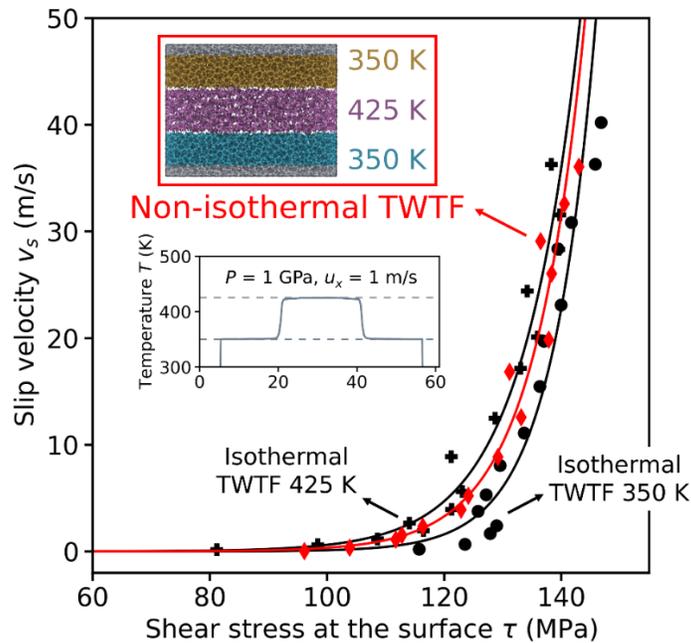

Figure 8 – Slip velocities as a function of the wall shear stress for the NI-TWTF system, in which the lubricant is thermalized to 425 K while the walls are kept at 350 K (solid red line and diamonds), compared to the isothermal TWTF cases at 350 and 425 K (solid black lines, circles

and "+" symbols). $P = 1$ GPa in all these simulations. The insets show schematically the target temperatures of the different regions of the NI-TWTF system and the temperature profile with $u_x$ = 1 m/s. Dashed lines in the temperature profile indicate the target temperatures of the thermostat.

**Building the TCM: Pressure- and temperature dependent viscosity law**

Another essential ingredient for the TCM is a temperature-dependent viscosity law. Introducing this law in the TCM would allow us to quantitatively evaluate the prevailing velocity accommodation mode, i.e., whether shearing occurs within the fluid or at the fluid/solid interface (slip). The dependence of the lubricant viscosity on temperature, pressure and shear rate can be modelled by the Eyring law for shear thinning in non-Newtonian fluids[38] with $T$- and $P$-dependent Eyring parameters:

$$\eta(\dot{\gamma}, T, P) = \frac{\tau_e(T,P)}{\dot{\gamma}} \sinh^{-1} \frac{\eta_N(T,P)\dot{\gamma}}{\tau_e(T,P)}. \tag{8}$$

The Eyring stress $\tau_e(T,P)$ and the Newtonian viscosity $\eta_N(T,P)$ are parametrized from MD simulations of the bulk fluid that will be described in a future publication[39]. The $P$-dependence of the Eyring stress $\tau_e$ is roughly linear:

$$\tau_e(T,P) = A_{\tau_e} P + B_{\tau_e}, \tag{9}$$

while there is hardly a $T$-dependence. The Newtonian viscosity $\eta_N$ is described by the Roelands equation[40]:

$$\eta_N(T,P) = \eta_0 e^{(\ln \eta_0 + 9.67) \cdot \left(-1 + \left(1 + \frac{P}{P_0}\right)^Z \cdot \left(\frac{T_0 - 138.15}{T - 138.15}\right)^S\right)}, \tag{10}$$

where $P_0 = 196.16$ MPa and $T_0 = 297.85$ K are reference pressure and temperature, respectively. The values of the fitting parameters are $A_{\tau_e} = 1.29 \cdot 10^{-2}$, $B_{\tau_e} = 18.15$ MPa, $\eta_0 =$ $8.13 \cdot 10^{-5}$ Pa·s, $Z = 0.25$ and $S = 0.75$. In Figure S8 in the SI, we show that the viscosities calculated for the parallel channel are compatible with the ones of the bulk lubricant. Deviations of the viscosities of the confined fluid from the bulk values can be interpreted as finite size effects[41], which could be included in a modified viscosity model. Since we aim at the extrapolation to wider channels using our TCM, we expect the bulk viscosity law to be adequate to study the competition between slip and viscosity. Note that viscosities calculated with the L-OPLS force field[16] are somewhat overestimated compared to the experimentally measured high-pressure viscosities of hydrocarbons[42] and therefore predictions of our TCM slightly overemphasize the role of slip.

**Combining wall slip, ITR and viscosity laws in the TCM**

To predict friction in a lubricated parallel channel of arbitrary lubricant film height $h$ and DLC film thicknesses $h_{\text{DLC}}$ beyond the scales of our MD simulations, we combine the slip, ITR and viscous laws in the TCM. The model is first validated with the MD results shown in the previous section and then used to investigate the conditions under which slip dominates over thermally induced viscosity reduction and vice versa. Our TCM model is briefly outlined below.

In a parallel channel (Figure 9), the shear stress at the lubricant/solid interface can be obtained by solving numerically the force balance between Eyring's equations for viscosity[38] and slip[28]:

$$\tau = \tau_e \sinh^{-1} \frac{\eta_N \dot{\gamma}}{\tau_e} = \tau_0 \cdot \sinh^{-1} \frac{v_s}{v_0}. \tag{11}$$

The parametrization of the Eyring stress $\tau_e(P,T)$ and the Newtonian viscosity $\eta_N(P,T)$ in the PAO4 viscosity model as wells as the Eyring stress $\tau_0(P,T)$ and Eyring velocity $v_0(P,T)$ in the PAO4/DLC slip model were the subject of the previous paragraphs.

The shear stress obtained by the numerical solution of Eq. (11) can be used to calculate the temperature profiles in the solid and the lubricant. For a given sliding velocity $u_x$, the friction power $\tau u_x$ released per unit area by the lubricant is conducted in equal parts in the upper and lower bodies. Therefore, the heat current into the lower DLC coating is $J_Q = -\frac{\tau u_x}{2}$. This can be inserted into Fourier's law $J_Q = -k_{\text{DLC}} \, dT/dz$ and, after one spatial integration with respect to $z$, the temperatures $T_{\text{surf}}$ of the DLC at the surface facing the lubricant is obtained:

$$T_{\text{surf}} = T_{\text{b}} + \frac{\tau u_x h_{\text{DLC}}}{2 k_{\text{DLC}}}. \qquad (12)$$

Here, $T_{\text{b}}$ is the substrate temperature of the DLC coating with thickness $h_{\text{DLC}}$ and thermal conductivity $k_{\text{DLC}}$. Because of the ITR, a temperature discontinuity is present at the DLC/PAO4 interface, located at $z = 0$:

$$T_{\text{lub}}(z=0) = T_{\text{surf}} + R_{\text{th}} \frac{\tau u_x}{2}. \qquad (13)$$

From the heat conduction equation for the lubricant $-k_{\text{PAO4}} \frac{d^2 T}{dz^2} = \tau \frac{dv(z)}{dz}$ and assuming a Couette velocity profile in the parallel channel, a parabolic temperature profile results from a twofold integration with respect to $z$:

$$T_{\text{lub}}(z) = T_{\text{lub}}(z=0) + \frac{\tau}{k_{\text{PAO4}}} \left( \frac{u_x}{2} - v_s \right) z \left( 1 - \frac{z}{h} \right). \qquad (14)$$

We note that Eqs. (12)-(14) can be easily adapted to describe an asperity contact between two infinite thermal half-spaces. This requires merely a replacement of $h_{\text{DLC}}$ with the radius $r_{\text{asp}}$ of the contact area (see the SI for details).

In our continuum model, Eq. (11) is initially solved for given $P$ and $u_x$ under isothermal conditions (i.e., $T(z) = T_b$). The resulting $\tau$ is used to solve Eqs. (12)-(14). In a second iteration, Eq. (11) is solved again with an updated global temperature equal to the average between the highest and lowest temperature in the fluid. This iterative problem is solved until self-consistency, when both the update on temperature and shear stress are below $1 \cdot 10^{-4}$ K and $1 \cdot 10^{-4}$ MPa, respectively.

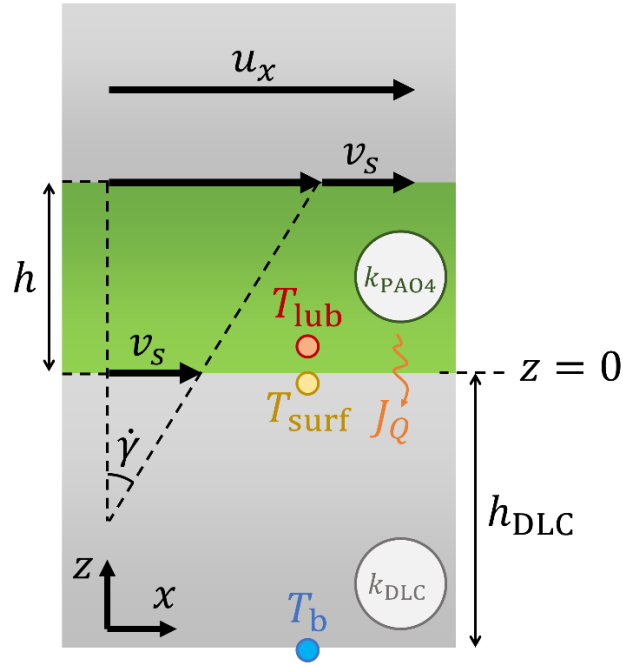

Figure 9 – Scheme of a lubricated parallel channel with slip and thermal dissipation. The top and bottom DLC slabs are equivalent. A Couette flow is established in the lubricant due to the constant sliding velocity $u_x$. Slip velocity $v_s$ affects the shear rate $\dot{\gamma}$. By setting the position of the bottom fluid/solid interface at $z = 0$, the top interface is at $z = h$. The thickness of the DLC slabs is $h_{\text{DLC}}$. The conductivities of the lubricant and the slabs are $k_{\text{PAO4}}$ and $k_{\text{DLC}}$, respectively. $T_b$, $T_{\text{surf}}$ and $T_{\text{lub}}$ are the temperature of the substrate far away from the lubricant, of the substrate in contact with the lubricant, and of the lubricant at an arbitrary value of $z$, respectively. $J_Q$ is the heat flux that, in case of viscous heating, flows from the lubricant to the substrates.

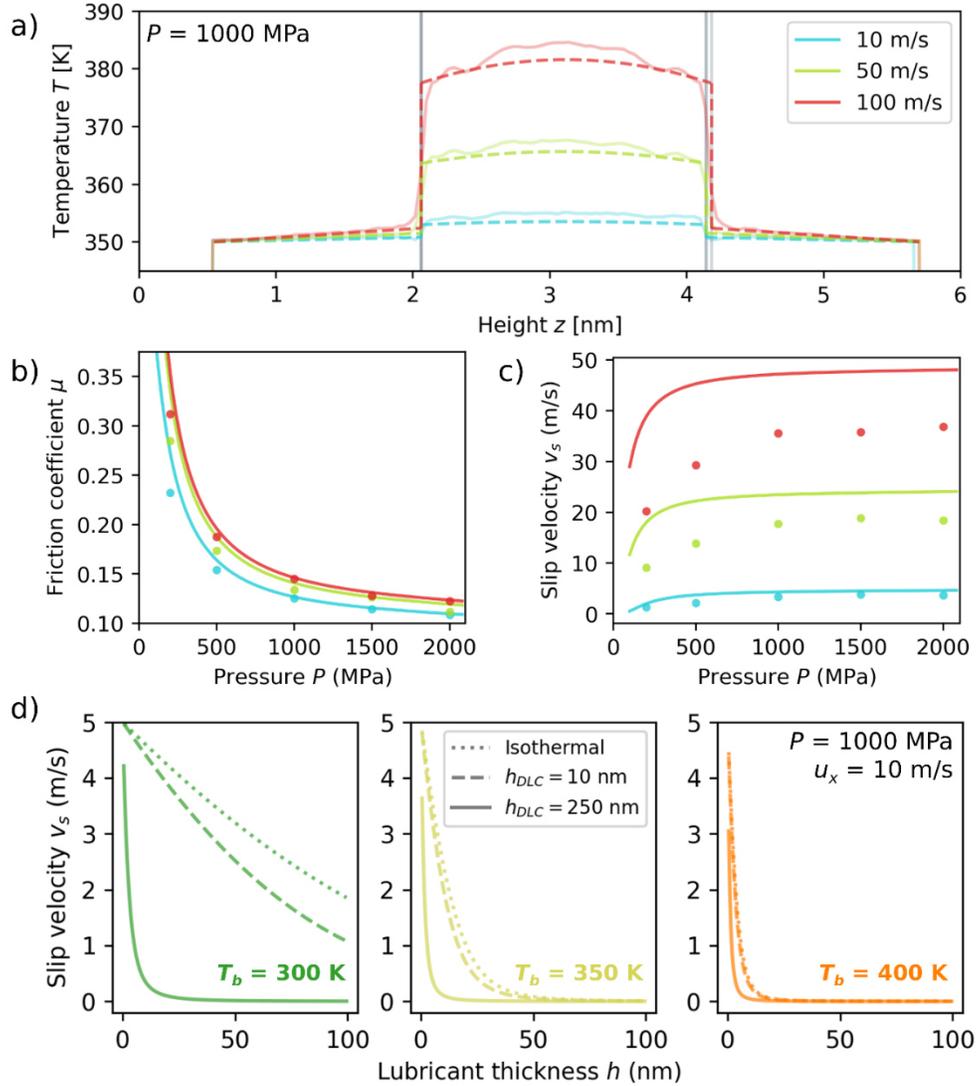

Figure 10 – a) Temperature profiles at $u_x$ = 10, 50 and 100 m/s and $P$ = 1 GPa predicted by the TCM (dashed lines) compared with the MD results (solid lines). b) Friction coefficient $\mu = \tau/P$ and c) slip velocity $v_s$ for the MD systems in a) (dots) compared to the model (solid lines). d) Slip velocity $v_s$ for different lubricant film thicknesses $h$ at $P$ = 1 GPa, $u_x$ = 10 m/s, and for different temperatures $T_b$ = 300, 350 and 400 K in the outermost regions of the a-C. The dotted lines show the results of the model without thermal effects (i.e., without any thermal barriers). The dashed and solid lines show the effect of viscous heating when 10 and 250 nm-thick a-C substrates are considered. The conductivities $k_{DLC}$ and $k_{PAO4}$ were obtained by fitting the temperature profiles

from the MD simulations with Eqs. (12) and (14). In panels a-c, for each sliding velocity, the conductivities were set as the average value at all pressures ($P$ = 0.2, 0.5, 1.0, 1.5 and 2.0 GPa). In panel d, they were set as the value at $P$ = 1 GPa and $u_x$ = 10 m/s, without noticeable changes from using the average value as described above. All the values of $k_{\text{DLC}}$ and $k_{\text{PAO4}}$ considered here are reported in Table S4.

Now we turn to the validation of the continuum model against the MD data of the TW system at $T_b = 350$ K and $P = 1$ GPa. Figure 10a displays the temperature profiles for three sliding velocities ($u_x = 10, 50$ and $100$ m/s). The TCM describes the MD profiles remarkably well. The same holds for the shear stress (see the friction coefficient $\mu = \tau/P$ in Figure 10b). Slip is also qualitatively reproduced (Figure 10c), and the agreement is even quantitative at low sliding velocity. The discrepancy at high sliding velocity in panel c (overestimation of $v_s$ by the TCM of around 50%) most likely results from inaccuracies in the prediction of the slip velocity at high $v_s$ values by our Eyring constitutive slip law.

Although the estimation of the slip velocity with the current continuum model is semi-quantitative, it is very instructive to discuss the predictions of the TCM for the system at $P$ = 1 GPa and $u_x$ = 10 m/s, for which the agreement with the MD is remarkably good. Figure 10d shows the slip velocity $v_s$ as a function of the lubricant film thickness $h$ for three different external temperatures $T_b$ (300, 350 and 400 K). For all temperatures, $v_s(h)$ decreases with an increase in $h$ for a given sliding velocity, since the shear rate also decreases with increasing $h$. As expected, the frictional response of thick enough confined lubricant films is governed by viscous shearing of the lubricant suggesting that no-slip boundary conditions are adequate for large enough $h$. Nevertheless, the rate at which $v_s$ decreases with $h$ depends on the temperature and the heat conductivities of the lubricant and the substrate. Three different cases are compared in Figure 10d. In the first one, viscous heating is not considered, and the temperature of the lubricant and the slabs is kept spatially constant at $T(z) = T_b$. This situation results in the slowest decay of $v_s(h)$ (see in the dotted lines in Figure 10d). When viscous heating is considered, the thickness of the substrate plays an important role in the heat dissipation. The dashed and solid lines show the slip velocity for DLC thicknesses $h_{DLC}$ of 10 and 250 nm, respectively. For a 10 nm-thick DLC film (a value still compatible with the size of our MD simulations), the decrease of $v_s(h)$ is similar to the $v_s(h)$ decay under isothermal conditions (compare dashed with dotted lines in Figure 10d). In this case, viscous heating has obviously only a minor influence on slip, since the DLC slabs are not insulating enough to cause essential viscous heating of the PAO4. However, for a DLC thickness $h_{DLC}$ of 250 nm (representative of coatings in tribological experiments and applications) the slip velocity becomes rapidly negligible. For instance, at $T_b = 300$ K (leftmost panel in Figure 10d), $v_s(h = 8.5$ nm$)$ is smaller than 10% of its maximum value at $h = 1$ nm, and drops to less than 1% at $h = 27.5$ nm. At higher initial temperatures $T_b = 350$ and 400 K (center and rightmost panel in Figure 10d) the slip velocity decreases even faster because the lubricant is less viscous than at $T_b = 300$ K.

As already mentioned above, these results can be also interpreted in the light of an asperity contact on two infinitely thick DLC coatings. For a radius $r_{asp} = 10$ nm of the contact area between the two DLC bodies, slip can play an essential role, especially for low temperatures (see dashed curves in Figure 10d, while larger radii lead to a strong reduction in slip velocity (see dotted curves in Figure 10d for $r_{asp} = 250$ nm). As surfaces in real tribological experiments are typically rough and asperity contact areas are usually micron-sized, the contribution of slip is expected to be negligible in the TEHL regime, while it remains dominant in the boundary lubrication regime. On the other hand, thin DLC coatings and sub-micron asperity contacts, especially if terminated by slippery graphitic structures, could drive a tribological system into a slip-dominated regime.

## Conclusions

Motivated by a long-standing debate in the literature regarding the role of temperature on wall slip in lubricated tribological contacts, we explored the slip properties of a 4 cSt polyalphaolefin base oil in a nanosized DLC channel. Our non-reactive MD study elucidated two aspects: (i) how the slip properties change with temperature and (ii) for which experimental conditions wall slip dominates over PAO4 shearing at elevated temperatures due to viscous heating. Firstly, we parametrized a temperature- and pressure-dependent slip law. For a given shear stress, higher temperatures lead to stronger slippage of the lubricant molecules on the surfaces, confirming the assumption that slip can be modeled as an activated process. Secondly, we parametrized a pressure-dependent law for the interfacial thermal resistance to describe heat transfer across the lubricant/solid interface. The ITR originating from viscous heating decreases exponentially with the pressure (and the shear stress) at the wall, indicating that thermal coupling between lubricant and surface increases with the increase of the interfacial stress. Interestingly, additional MD simulations with an isothermal lubricant in contact with cool isothermal walls suggest that the temperature of the surface plays a negligible role on slip, and it is the lubricant temperature at the surface that determines the magnitude of slip velocity.

Based on the *T*- and *P*-dependent slip and viscosity laws and the *P*-dependent ITR law, we developed a simple non-isothermal continuum model for a lubricant subject to viscous heating in a tribological contact formed by two parallel walls. The model indicates that slip only plays a role in the boundary lubrication regime, while viscosity reduction due to viscous heating dominate otherwise. While providing valuable insights on which conditions make slip dominate over lubricant shearing, the model still does not satisfactorily describe roughness nor confinement effects. For instance, the lubricant rheology in nanochannels deviates from the one of the bulk fluid[41]. Future research will focus on studying the connection between slip and the ITR, the influence of surface roughness as well as other substrates with higher heat conductivities than the one of DLC. Nevertheless, these results represent a crucial step towards a MD-informed thermo-elasto-hydrodynamic lubrication model to accurately describe lubricant flow from the extreme boundary up to the hydrodynamic regime, enabling a smart design of lubricated interfaces in which the friction performance can be chosen *ad hoc* based on thermal properties of the contact. By designing boundary-lubricated contacts where the thickness of the top and bottom insulating substrates differs, lubricant flow, slip and, therefore, the friction performance of the systems can be tuned selectively based on the heat transport properties of the individual fluid/substrate interfaces.

ASSOCIATED CONTENT

**Supporting Information**. Details on the fitting procedure of the slip law, slip velocities in the TWTF and TW systems with error bars, dependence of the ITR on temperature, viscosities in the parallel channel and in the bulk fluid, derivation of the thermal laws, conductivities of PAO4 and DLC in the TCM, temperature profiles in the TW systems, temperature of the lubricant and slip velocity as a function of the sliding velocity in the TW systems (PDF); Summary of the results of the MD simulations (CSV).

**Data availability statement.** Input, representative output and selected auxiliary files will be uploaded on Zenodo upon acceptance.

AUTHOR INFORMATION

**Corresponding Authors**

E-mail addresses:

*S.P.: stefan.peeters@iwm.fraunhofer.de, **M.M.: michael.moseler@iwm.fraunhofer.de

**Author Contributions**

S.P.: Conceptualization, investigation, formal analysis, software, visualization, writing – original draft, review & editing. E.J.G.: Formal analysis, writing – review & editing. F.S.: Investigation. T.R., K.F. and G.M.: Supervision, writing – review & editing. M.M.: Conceptualization, formal analysis, validation, supervision, project administration, writing – original draft, review & editing.

**Funding Sources**

This project has received funding from the European Research Council (ERC) under the European Union's Horizon 2020 research and innovation programme (ERC Advanced Grant LubeTwin, agreement No. 101201061). We also acknowledge financial support by the German Research Foundation (Deutsche Forschungsgemeinschaft, DFG) within the Research Unit 5099 "Reducing complexity of nonequilibrium systems". S.P. and M.M. additionally acknowledge funding from the Deutsche Forschungsgemeinschaft (DFG) via project 556195707 (MO 879/31-1).

ACKNOWLEDGMENT

The authors gratefully acknowledge the Gauss Centre for Supercomputing e.V. (www.gauss-centre.eu) for funding this project by providing computing time through the John von Neumann Institute for Computing (NIC) on the GCS Supercomputer JUWELS at Jülich Supercomputing Centre (JSC), as well as the high-performance Compute Cluster "HoreKa" at the NHR Center

NHR@KIT, which is jointly supported by the Federal Ministry of Education and Research and the state governments participating in the NHR (www.nhr-verein.de/unsere-partner).

ABBREVIATIONS

a-C, amorphous carbon; PAO4, 4 cSt polyalphaolefin; EHL, elasto-hydrodynamic lubrication; TEHL, thermo-elasto-hydrodynamic lubrication; RLE, Reynolds lubrication equation; DLC, diamond-like carbon; MD, molecular dynamics; TWTF, thermostat walls thermostat fluid; NI-TWTF, non-isothermal thermostat walls thermostat fluid; TW, thermostat walls; ITR, interfacial thermal resistance; TCM, thermal continuum model; PPPM, particle-particle particle-mesh; EV, electric vehicles.

# When wall slip wins over shear flow: A temperature-dependent Eyring slip law and a thermal multiscale model for diamond-like carbon lubricated by a polyalphaolefin oil

Stefan Peeters[*a,b], Edder J. García[a,b], Franziska Stief[a,c], Thomas Reichenbach[a], Kerstin Falk[a], Gianpietro Moras[a], Michael Moseler[**a,b,c]

[a]*Fraunhofer IWM, MikroTribologie Centrum µTC, Wöhlerstraße 11, 79108 Freiburg, Germany*

[b]*Freiburg Materials Research Center, University of Freiburg, Stefan-Meier-Straße 21, 79104 Freiburg, Germany*

[c]*Institute of Physics, University of Freiburg, Herrmann-Herder-Straße 3, 79104 Freiburg, Germany*

Corresponding authors:
[*]stefan.peeters@iwm.fraunhofer.de
[**]michael.moseler@iwm.fraunhofer.de

# S1 – Details on the fitting procedure of the slip law

The simplest way is to fit Eq. 1 without any constraint. A set of $\tau_0$ and $v_0$ is directly obtained. This procedure was adopted for the Eyring curves shown in Figures 3 and 5. The pressure dependence of $\tau_0$ and $v_0$ can be fitted via:

$$\tau_0(P) = A_{\tau_0} P + B_{\tau_0} \qquad (S1)$$

and

$$v_0(P) = A_{v_0} e^{B_{v_0} P + C_{v_0} P^4}, \qquad (S2)$$

where $A_{\tau_0}, B_{\tau_0}, A_{v_0}, B_{v_0}, C_{v_0}$ are fit constants, in agreement with our previous work[1]. Figure S1 show three exemplary plots of $\tau_0(P)$ and $v_0(P)$ at $T$ = 300, 350 and 400 K, while Table S1 reports the parameters obtained by the unconstrained fitting of Eqs. S1 and S2 for different systems considered in this work. Finally, Table S2 includes the fitting parameters of Eq. 1 for the systems shown in Figure 8 in the main text.

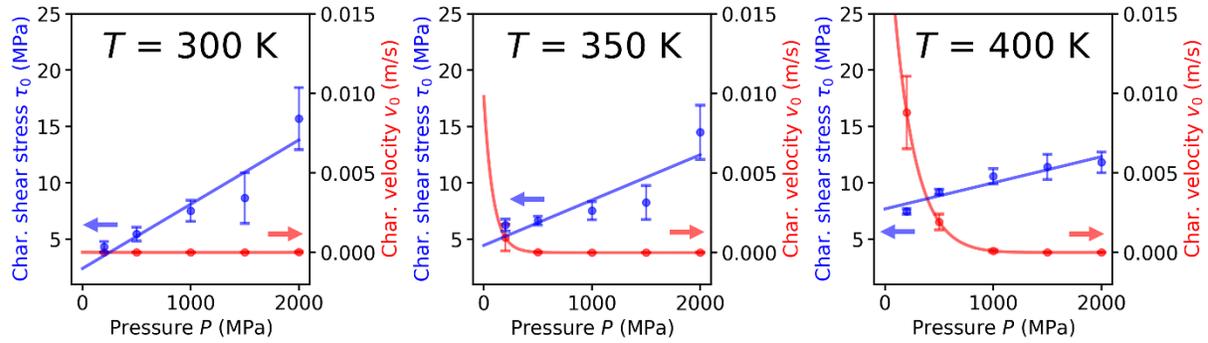

**Figure S1.** Dependence of the characteristic parameters of the slip laws on pressure at 300, 350 and 400 K and the corresponding fits to Eqs. S1 and S2. Here, 1 m/s ≤ $u_x$ ≤ 100 m/s. The error bars represent the standard error of the estimated parameters from the free ODR fit.

**Table S1.** Fitting parameters of Eqs. S1 and S2 for the different simulation systems (1 m/s ≤ $u_x$ ≤ 100 m/s).

| T (K) | Setup | $A_{\tau_0}$ (×10$^{-3}$) | $B_{\tau_0}$ (MPa) | $A_{v_0}$ (m/s) | $B_{v_0}$ (GPa$^{-1}$) | $C_{v_0}$ (GPa$^{-4}$) |
|---|---|---|---|---|---|---|
| 250 | TWTF | 4.05 ± 1.26 | 4.42 ± 1.55 | -6.98 ± 2.13 | -10.4 ± 10.6 | 0.91 ± 1.35 |
| 300 | TWTF | 5.71 ± 1.24 | 2.40 ± 1.53 | -11.3 ± 0.1 | -5.05 ± 0.46 | 0.64 ± 0.05 |
| 350 | TWTF | 4.02 ± 1.30 | 4.46 ± 1.60 | -4.61 ± 0.00 | -11.7 ± 0.0 | 0.98 ± 0.00 |
| 350 | TW | 3.73 ± 0.28 | 5.35 ± 0.35 | -4.59 ± 0.02 | -8.25 ± 0.11 | 0.50 ± 0.17 |
| 400 | TWTF | 2.31 ± 0.44 | 7.68 ± 0.54 | -3.74 ± 0.00 | -4.95 ± 0.01 | -0.59 ± 0.03 |
| 450 | TWTF | 2.41 ± 0.48 | 8.41 ± 0.59 | -0.61 ± 0.04 | -10.5 ± 0.2 | 0.67 ± 0.49 |

**Table S2.** Characteristic parameters of Eq. 1 for selected systems, including those described in Figure 8.

| Setup | $\tau_0$ (MPa) | $v_0$ (m/s) |
|---|---|---|
| Isothermal TWTF, 350 K | 9.70 ± 0.95 | (2.18 ± 3.12) ×10$^{-5}$ |
| Non-isothermal TWTF | 11.39 ± 0.96 | (2.46 ± 2.54) ×10$^{-4}$ |
| Isothermal TWTF, 425 K | 11.75 ± 0.84 | (4.13 ± 3.45) ×10$^{-4}$ |
| Isothermal TWTF, 450 K | 12.77 ± 0.54 | (1.23 ± 0.56) ×10$^{-3}$ |
| TW, 350 K | 13.01 ± 1.58 | (9.61 ± 12.65) ×10$^{-4}$ |

Nevertheless, $\tau_0$ and $v_0$ modify the shape and the position of the slip curve in a similar way, and this issue introduces large uncertainties on the fitted $\tau_0$ and $v_0$. In case an additional law can be used to constrain the values of $\tau_0$ (or $v_0$), a subsequent fit could refine $v_0$ (or $\tau_0$), offering higher accuracy than the free fit. Such a procedure was used to derive the *T*- and *P*-dependent slip model shown in Figure 4 by using the Eyring law for $\tau_0(T)$ and $v_0(T)$, and could also be adopted for the results shown in Figure 5 by using the empirical law of $\tau_0(P)$ in Eq. S1. The details of these two fitting procedures are presented in the following, while Figure S2 schematically shows the different strategies adopted to fit the slip laws in this work.

**Constrained fit for $\tau_0(P)$ (Figure S2b)** – Although this procedure is not shown in the main text, we describe a possible strategy to refine $\tau_0$ or $v_0$ even without introducing temperature dependence. As a first step, the MD data is fitted with Eq. 1 without constraints. A first set of $\tau_0$ and $v_0$ is obtained for each pressure. The linear fit of $\tau_0(P)$ (Eq. S1) is quite satisfactory, and the values of $\tau_0$ can be recalculated for each pressure so that they lie on the fitted line. A subsequent fit of Eq. 1 can be performed to obtain a new set of values for $v_0$ by keeping $\tau_0$ fixed. This significantly reduces the error for the estimation of $v_0$. The choice of fixing the values of $\tau_0$, so that they lie on $\tau_0(P)$, is arbitrary. Indeed, the best fit is often obtained without constraints. However, the slip curves change only slightly by introducing this constrained fit (typically with an almost negligible increase of the slip velocity at high shear stress, see Figure S3).

**Constrained fit for $\tau_0(T)$ and $v_0(T)$** (Figure S2c) – A similar strategy can be adopted for the temperature dependence of the characteristic parameters. Here, the Eyring laws from Eqs. 3 and 4 can be used. To additionally include the pressure dependence, the parameters $c_1$, $c_2$ and $c_3$ from those two equations should be evaluated as a function of pressure. A linear fit (Eq. 5) is surprisingly effective to describe $c_1(P)$, and we follow a similar procedure as before, with the values of $c_1$ being recalculated for each pressure to match the prediction of the fitted $c_1(P)$. Then, these parameters are used to recalculate $\tau_0$ for each temperature and pressure. These new values of $\tau_0$ are kept fixed for a constrained fit of Eq. 1 to obtain $v_0$. This represents the beginning of a second iteration to parametrize the slip model. This time, $v_0(T)$ is fitted with Eq. 4, yielding $c_2$ and $c_3$ for each pressure. A 3$^{rd}$-degree polynomial fit is then chosen for both $c_2(P)$ and $c_3(P)$ (Eqs. 6 and 7). The values of $c_2$ and $c_3$ are recalculated to lie on the respective curves, and from each pair, new $v_0$ are obtained for each temperature and pressure. By fitting Eq. 1 again, a third iteration begins, this time with fixed $v_0$ and free $\tau_0$. The accuracy of the new $\tau_0$ already improves compared to the first free fit, yet it is convenient to further refine the characteristic parameters to increase the predictive power of the slip model. For this reason, we follow the same procedure as before: $c_1(P)$ is fitted with Eq. 5, providing two parameters of the slip model, then the values of $c_1$ and therefore $\tau_0$ are updated to match the recalculated dependencies, and a constrained fit of Eq. 1 provides a final set of $v_0$. Another polynomial fit of $c_2(P)$ and $c_3(P)$ determines the remaining eight parameters of the slip model. It should be evident that, unlike the previous strategy, introducing both temperature and pressure dependence of the slip parameters requires a double step when passing information from the MD data to the model and vice versa. We also verified that including an additional loop does not significantly improve the quality of the fit.

Unconstrained fits are consistently used in the main text for two reasons: (i) no assumption based on empirical laws is required on $\tau_0$ and $v_0$ and (ii) the temperature and pressure dependence of slip is not available for every simulation setup. The results with a constrained fit are generally shown in the Supporting Information for illustrative purposes, while this procedure is essential in Figure 4 to enable a good agreement with the Eyring law (see Figure S4 for $\tau_0(T)$ and $v_0(T)$ resulting from an unconstrained fit).

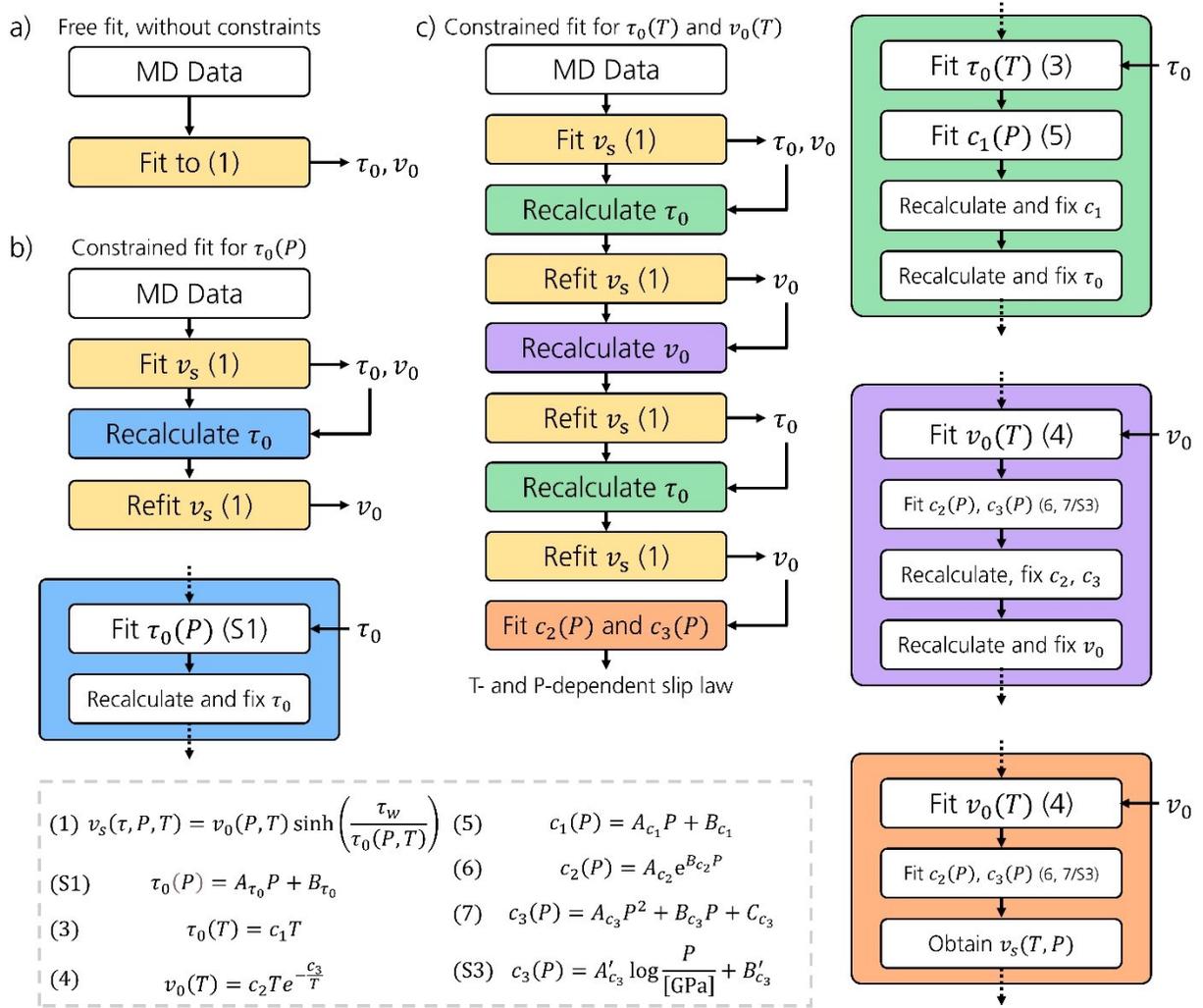

**Figure S2.** Schemes for the different fitting procedures adopted in this work. The details of the blue, green, purple and orange boxes are shown below and on the side of the main paths in b and c, respectively. The numbering of the equations is the same as in the main text and the SI.

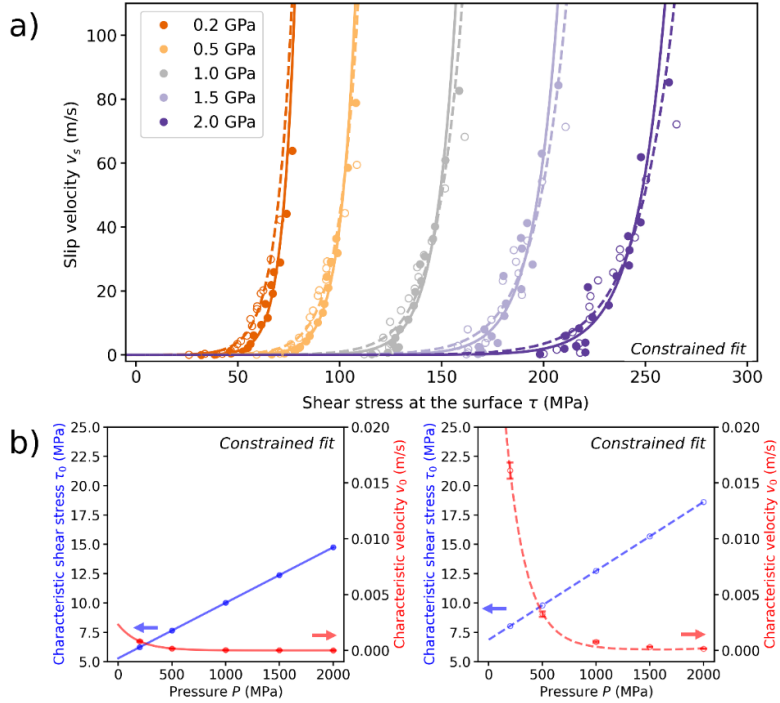

**Figure S3.** (a) Slip velocities as a function of the shear stress for all the simulations at 350 K (1 m/s $\leq u_x \leq$ 200 m/s) and the corresponding slip laws fitted to Eq. 1 using a constrained fit for $\tau_0(P)$. The curves are almost indistinguishable to the ones shown in Figure 5 in the main text. (b) Dependence of the characteristic parameters of the slip laws on pressure and the corresponding fit to Eqs. S1 and S2 using the constrained fit. The error bars represent the standard error of the estimated parameters from the ODR fit. Symbols and lines in (b) follow the same convention as in (a).

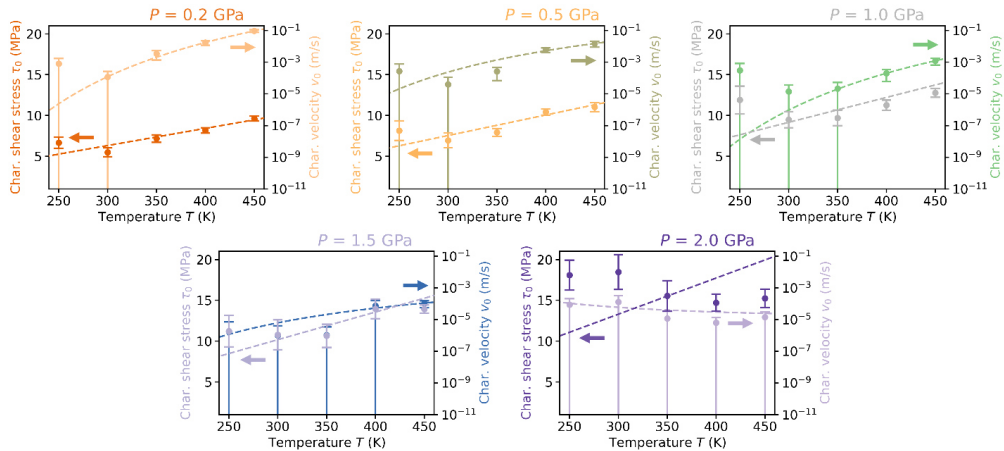

**Figure S4.** $\tau_0(T)$ and $v_0(T)$ fitted by Eqs. 3 and 4 after an unconstrained fit for different pressures. Here, 1 m/s $\leq u_x \leq$ 200 m/s, yet the results are equivalent if 1 m/s $\leq u_x \leq$ 100 m/s.

***T*- and *P*-dependent slip model with a logarithmic dependence of *c₃(P)*** – We provide an additional slip model where $c_3(P)$ is fitted with a logarithmic function to ensure a monotonic increase of the activation temperature:

$$c_3(P) = A'_{c_3} \log \frac{P}{P_0} + B'_{c_3}, \tag{S3}$$

where $P_0 = 1$ GPa, since the fitting was performed with pressures in GPa. The table below reports the fitting parameters converted for pressures expressed in MPa, as in Table 1 in the main text.

**Table S3.** Parameters obtained from the final fit of Eqs. 5, 6 and S3. The error on $B'_{c_3}$ was calculated as $\sigma_{B'_{c_3}(\text{MPa})} = \sqrt{\sigma_{B'_{c_3}(\text{GPa})}^2 + \left(\sigma_{A'_{c_3}(\text{GPa})} \log 10^{-3}\right)^2 + 2\sigma_{A'_{c_3}(\text{GPa})B'_{c_3}(\text{GPa})}}$, where $\sigma_k$ is the standard deviation of parameter $k$, the subscript (MPa/GPa) indicates the units of pressure to be used in Eq. S3, $\sigma_{A'_{c_3}(\text{GPa})B'_{c_3}(\text{GPa})}$ is the covariance of the parameters $A'_{c_3}$ and $B'_{c_3}$ for pressures in GPa.

| Parameter | Value | Parameter | Value |
|---|---|---|---|
| $A'_{c_1}$ | $9.6409 \cdot 10^{-6} \frac{1}{\text{K}}$ | $B'_{c_2}$ | $(-2.0616 \pm 0.2189) \cdot 10^{-3} \frac{1}{\text{MPa}}$ |
| $B'_{c_1}$ | $1.6070 \cdot 10^{-2} \frac{\text{MPa}}{\text{K}}$ | $A'_{c_3}$ | $(7.5200 \pm 0.5839) \cdot 10^2$ K |
| $A'_{c_2}$ | $(15.246 \pm 3.332) \frac{\text{m}}{\text{s} \cdot \text{K}}$ | $B'_{c_3}$ | $(1.2909 \pm 0.4009) \cdot 10^3$ K |

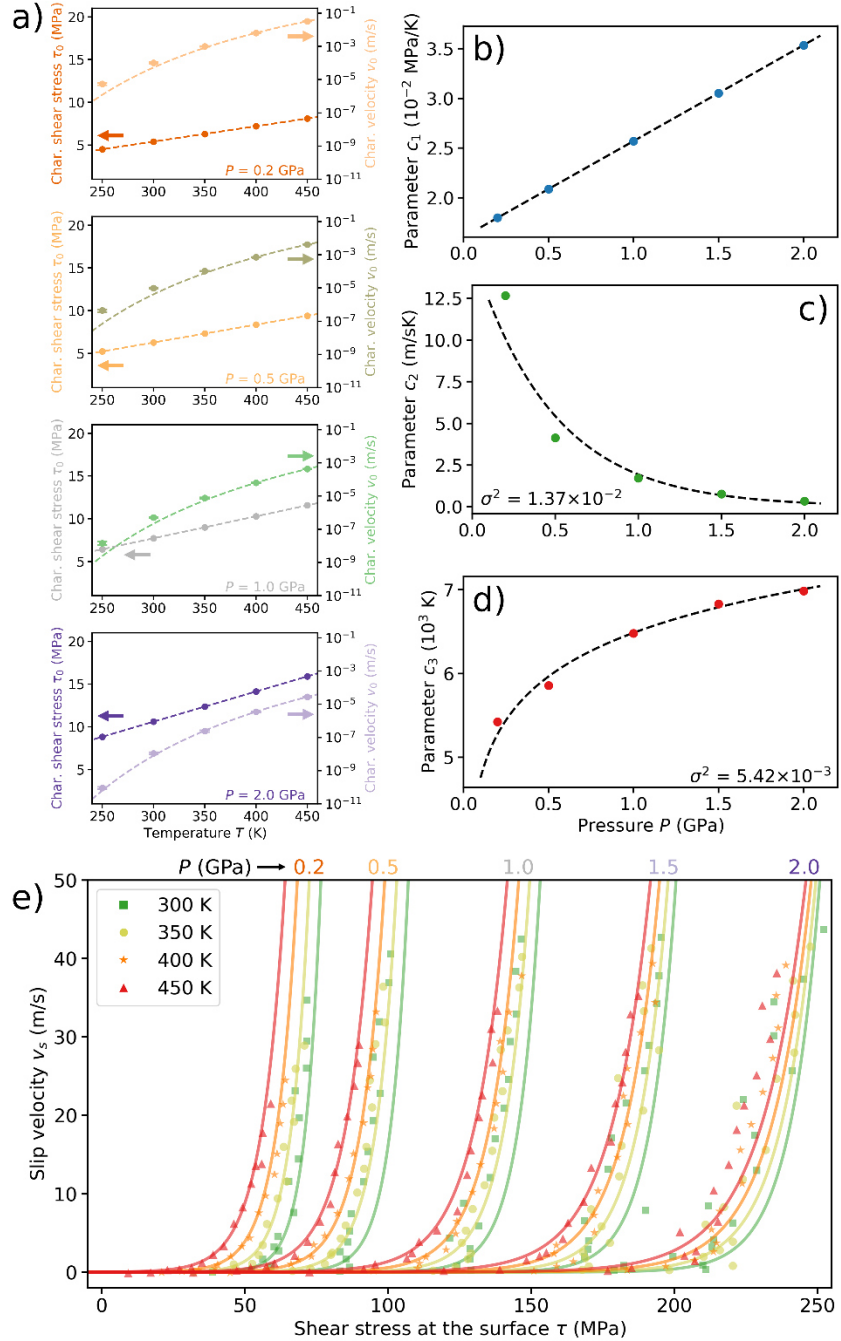

**Figure S5.** (a) $\tau_0(T)$ and $v_0(T)$ fitted by Eqs. 3 and 4 after three iterations of the constrained fit in which $c_3(P)$ is fitted by Eq. S3 ($P$ = 1.5 GPa is omitted, 1 m/s $\leq u_x \leq$ 100 m/s). Barely visible error bars on $v_0$ represent the uncertainty of the estimation of these parameters after three iterations of the constrained fit. (b)-(d) show the respective fit of $c_1(P)$, $c_2(P)$ and $c_3(P)$ with Eqs. 5, 6 and S3 after three iterations. $\sigma^2$ represents the residual variance of the corresponding ODR fits. (e) MD results of selected systems and the slip laws at the corresponding temperatures and pressures predicted by the resulting model.

# S2 – Slip velocities in the TWTF and TW systems with error bars

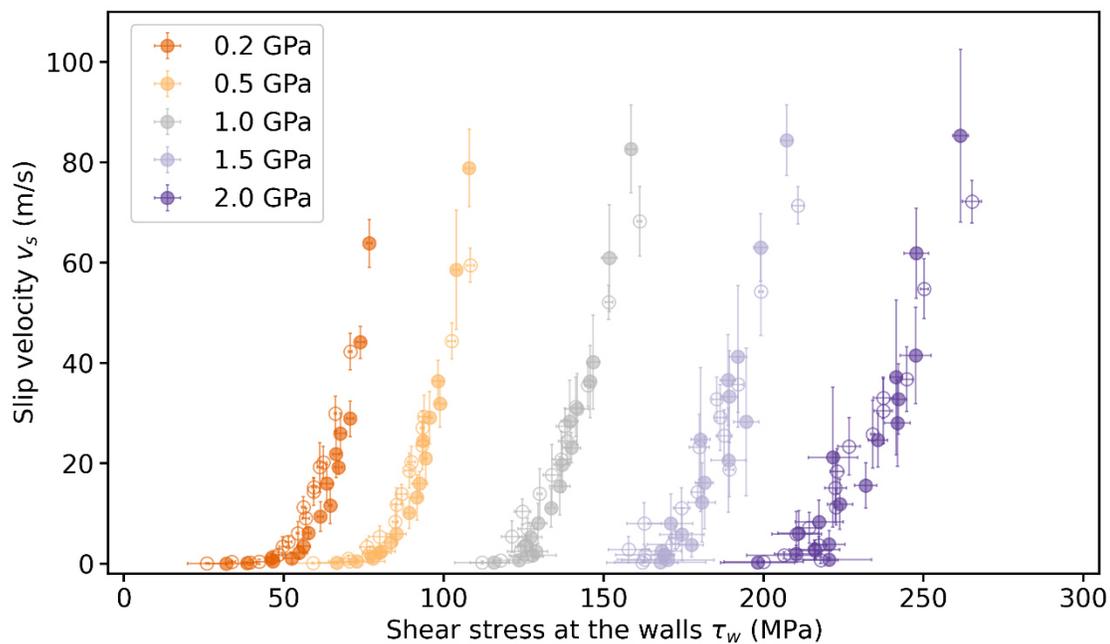

**Figure S6.** Slip velocities as a function of the shear stress at the surface for all the TWTF and TW simulations at 350 K (1 m/s ≤ $u_x$ ≤ 200 m/s). Filled and empty circles refer to the simulations with the TWTF and the TW schemes, respectively. The error bars represent the standard deviation of 5 independent evaluations, one for each nanosecond of the run. The Eyring fits are omitted for clarity.

# S3 – Dependence of the ITR on temperature

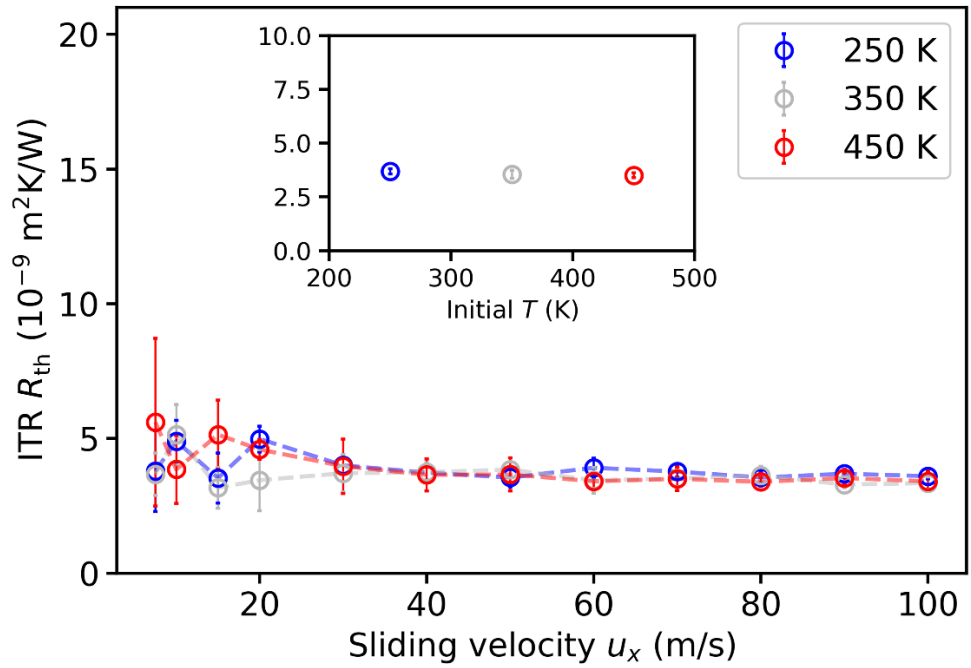

**Figure S7.** Interfacial thermal resistance ITR of the TW systems as a function of sliding velocity and initial temperature. The data points between $u_x$ = 1 and 2.5 m/s were discarded due to the large uncertainty of the resulting ITR. The error bars represent the standard deviation of 5 individual evaluations for each system. The dashed lines are a guide to the eye. The error bars in the inset mark the standard deviation of the ITR averaged in the range 40-100 m/s.

# S4 – Viscosities in the parallel channel and in the bulk fluid

Figure S8. a shows the shear-thinning behavior of the isothermal systems at 350 and 450 K, as well as the TW systems, compared to the same data for the NI-TWTF system. The viscosity of the fluid in contact with the surfaces was calculated as $\eta = \frac{\tau}{\dot{\gamma}}$. The shear-thinning curves were obtained by plotting $\eta(\dot{\gamma})$, and the data points were fitted with the Eyring law for bulk viscosity[2]:

$$\eta(\dot{\gamma}) = \frac{\tau_e}{\dot{\gamma}} \sinh^{-1} \frac{\eta_N \dot{\gamma}}{\tau_e}, \quad (S4)$$

where the fitting parameters $\tau_e$ and $\eta_N$ are the Eyring stress and the Newtonian viscosity. While the decrease in viscosity at constant shear rate can be observed by increasing the temperature from 350 to 450 K, the non-Newtonian behavior of the lubricant plays a dominant role. The curve for the NI-TWTF system lies between the ones at 350 and 450 K, while the data for the TW system, initially close to the ones of TWTF at 350 K, reach overall the lowest viscosity. The values of shear rate and viscosity for these systems are fully compatible with the ones obtained in the shearing simulations of the bulk fluid under similar thermodynamic conditions, as shown in panel b. The simulations of the bulk 1-decene trimer ($C_{30}H_{62}$), modeled by the L-OPLS potential[3], were performed using LAMMPS[4]. The volume of the simulation cell was filled with 260 molecules and was equilibrated for 17.5 ns at different combinations of temperature and pressures $T$ = 333, 373, 393 K and $P$ = 0.03, 0.25, 0.5, 1, 2 GPa, resulting in cell volumes between 4.6×5.1×7.4 nm³ and 5.2×5.7×8.3 nm³. A Nosé-Hoover thermostat with a time constant of 10 fs was used for temperature control. After equilibration, the simulations were run for 5 ns. Periodic boundary conditions were adopted in all dimensions using the SLLOD method[5] to apply shearing to the bulk volume every 0.5 fs. Shear velocities between 1 and 600 m/s were applied, leading to shear rates between $10^{-1}$ and $10^2$ ns$^{-1}$. All curves show the shear-thinning regime of $C_{30}H_{62}$. Higher viscosities are observed at higher pressures and temperatures, whereby the pressure has a stronger impact on viscosity. In general, the differences in viscosity decrease by increasing the shear rate.

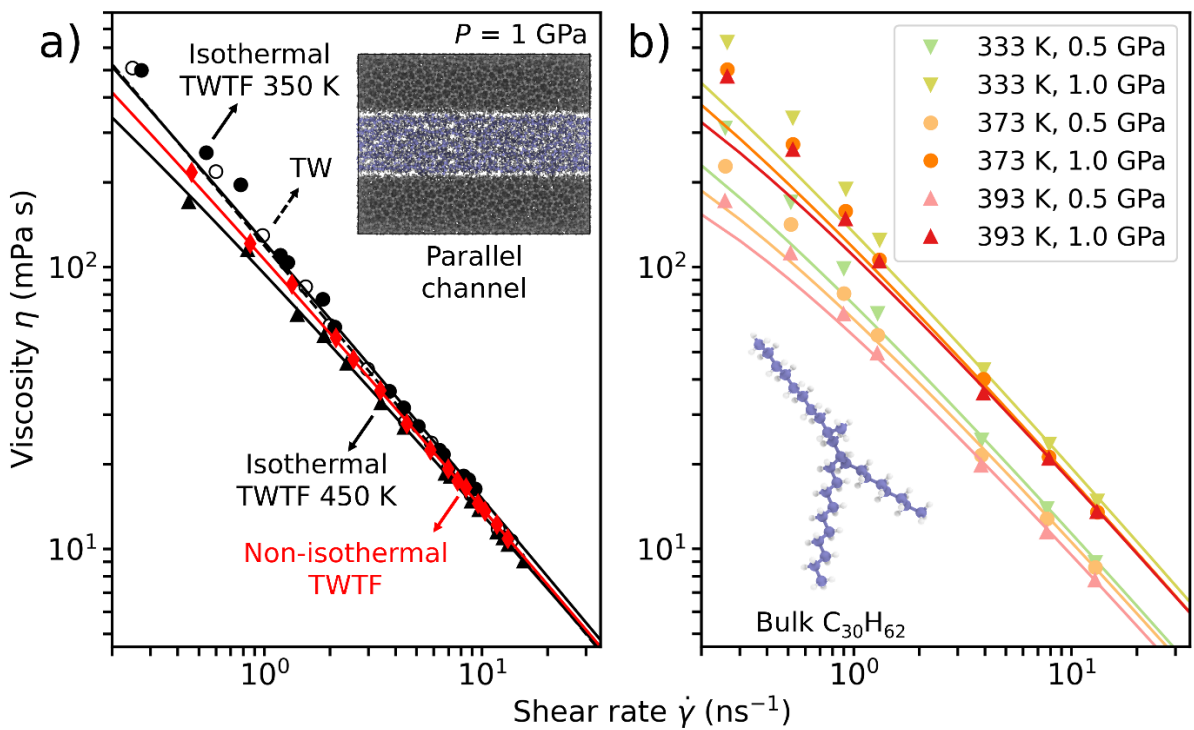

**Figure S8.** Shear-thinning behavior of the lubricant in the parallel channel (a), following the same convention regarding symbols, colors, and line styles as in Figure 8 in the main text (1 m/s ≤ $u_x$ ≤ 100 m/s), and of the bulk fluid (b). The lines represent the ODR fit of the datapoints with Eq. S4.

# S5 – Derivation of the thermal laws

In the model shown in Figure 9 in the main text, the bottom and top materials are equivalent (thermal conductivity $k_{\text{DLC}}$, thickness $h_{\text{DLC}}$, in contact with a thermal reservoir at temperature $T_{\text{b}}$). The heat current $J_Q$ from the lubricant to the solid material, according to Fourier's law, is

$$J_Q = -k_{\text{DLC}} \frac{dT}{dz} \tag{S5}$$

The heat conduction equation on the solid is

$$-k_{\text{DLC}} \frac{d^2T}{dz^2} = 0 \tag{S6}$$

since there are no heat sources within the sliding bodies. Integrating twice with respect to $z$ shows that a linear temperature profile is established on the solids:

$$\frac{dT}{dz} = \frac{dT}{dz}\bigg|_0 = -\frac{1}{k_{\text{DLC}}} J_Q(0) \rightarrow T(z) = T(0) - \frac{1}{k_{\text{DLC}}} J_Q(0) z \tag{S7}$$

$$T(-h_{\text{DLC}}) - T(0) = \frac{1}{k_{\text{DLC}}} J_Q(0) h_{\text{DLC}} \tag{S8}$$

$$J_Q = -k_{\text{DLC}} \frac{T_{\text{surf}} - T_{\text{b}}}{h_{\text{DLC}}} \tag{S9}$$

where $T_{\text{surf}}$ is the temperature of the solid at the surface. By knowing the heat current $J_Q$, the temperature of the surface is

$$T_{\text{surf}} = T_{\text{b}} - \frac{J_Q h_{\text{DLC}}}{k_{\text{DLC}}} \tag{S10}$$

Assuming that the lubricant slipping over the solid absorbs all the frictional heat followed by heat transport into the solid, the heat current into the lower body is given by the slip and the viscous contribution

$$|J_Q| = \tau v_s + \int_0^{\frac{h}{2}} \tau \dot{\gamma}\, dz = \tau \left( v_s + \int_0^{\frac{h}{2}} \frac{dv}{dz}\, dz \right) = \tau \left[ v_s + v\left(\frac{h}{2}\right) - v_s \right] = \tau \frac{u_x}{2} \tag{S11}$$

Therefore, the temperature at the surface of the solid is given by

$$T_{\text{surf}} = T_{\text{b}} + \frac{\tau u_x h_{\text{DLC}}}{2 k_{\text{DLC}}} \tag{12}$$

Let us consider the balance of the Eyring laws for viscosity and slip,

$$\tau = \tau_e \sinh^{-1} \frac{\eta_N \dot{\gamma}}{\tau_e} = \tau_0 \cdot \sinh^{-1}\left(\frac{v_s}{v_0}\right), \text{ with } \dot{\gamma} = \frac{u_x - 2 v_s}{h} \tag{S12}$$

where the parameters $\tau_0$, $v_0$ must be evaluated at the temperature of the first lubricant layer $T_{\text{lub}}(z=0)$, while the Eyring stress $\tau_e$ and the Newtonian viscosity $\eta_N$ can vary with temperature and, therefore, on the vertical position. We can derive the heat conduction equation in the lubricant:

$$-k_l \frac{d^2 T}{dz^2} = \tau \frac{dv(z)}{dz} + \tau v_s \delta(z) \tag{S13}$$

First, let us integrate from $z' = \frac{h}{2}$ to $z$:

$$-k_{\text{PAO4}} \int_{h/2}^{z} dz' \frac{d^2 T}{dz'^2} = \int_{h/2}^{z} dz' \left[\tau \frac{dv(z')}{dz'}\right]$$

$$-k_{\text{PAO4}} \left( \frac{dT}{dz}\bigg|_z - \frac{dT}{dz}\bigg|_{z=\frac{h}{2}} \right) = \tau \left( v(z) - v\left(\frac{h}{2}\right) \right) \tag{S14}$$

$$-k_{\text{PAO4}} \left( \frac{dT(z)}{dz} \right) = \tau \left( v(z) - \frac{u_x}{2} \right)$$

$$\frac{dT}{dz} = \frac{\tau}{k_{\text{PAO4}}} \left( \frac{u_x}{2} - v(z) \right)$$

Let us consider a Couette profile, in agreement with our MD results:

$$v(z) = v_s + \dot{\gamma} z = v_s + \frac{u_x - 2 v_s}{h} z \tag{S15}$$

$$\frac{dT}{dz} = \frac{\tau}{k_{PAO4}} \left( \frac{u_x}{2} - v_s - \frac{u_x - 2v_s}{h} z \right) \tag{S16}$$

Another integration yields:

$$T(z) - T(0) = \frac{\tau}{k_{PAO4}} \left( \left(\frac{u_x}{2} - v_s\right) z - \frac{\frac{u_x}{2} - v_s}{h} z^2 \right) \tag{14}$$

$$T(z) - T(0) = \frac{\tau}{k_{PAO4}} \left( \frac{u_x}{2} - v_s \right) z \left( 1 - \frac{z}{h} \right)$$

Next, we calculate the temperature jump between surface and lubricant at the interface. The interfacial thermal resistance is:

$$R_{th} = \frac{\Delta T}{J_Q} \tag{2}$$

yielding a temperature jump between solid and liquid equal to

$$T_{lub}(0) - T_{surf} = R_{th} J_Q = R_{th} \frac{\tau u_x}{2}. \tag{13}$$

Interestingly, there is a version of the TCM for the lubricated contact of an infinite solid surface and an asperity supported by another infinite solid surface (see Figure S9). Assuming, that the heat flux $J_Q = -\frac{\tau u_x}{2}$ passes through a circular contact area with radius $r_{asp}$ and that $u_x$ is small, the temperature on the surface is approximately given by Carslaw and Jaeger[6]:

$$T_{surf} - T_b \approx \frac{\tau u_x}{2} \cdot \frac{r_{asp}}{k_{DLC}}. \tag{S17}$$

A comparison with Eq. 12 reveals that the TCM formalism also holds for the situation described in Figure S9, provided $h_{DLC}$ is replaced by $r_{asp}$. Thus, the trends observed in the TCM with increasing thickness of the DLC layers are identical to the trends with increasing radius of a circular surface spot radiating heat into two infinitely thick DLC layers.

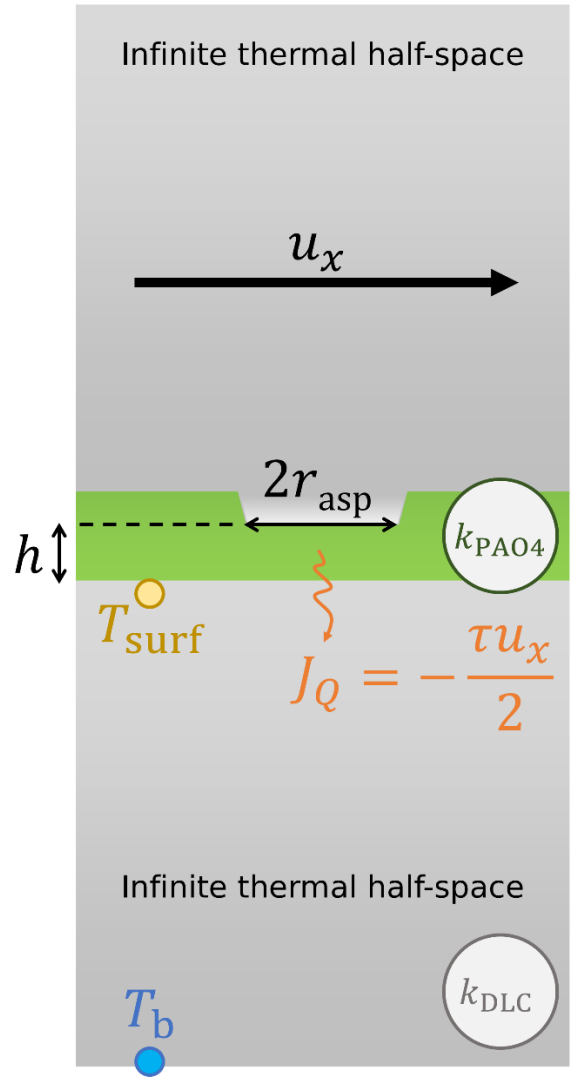

**Figure S9.** Scheme of a lubricated parallel channel between an infinite solid surface and an asperity supported by another infinite solid surface. Both surfaces are modelled as homogeneous thermal half spaces with heat thermal conductivity $k_{\text{DLC}}$. The sliding velocity $u_x$ is assumed to be small enough such that surface temperature $T_{\text{surf}}$ can be estimated from the solution of the heat conduction problem in a thermal half-space subject to a stationary heat flux $J_Q = -\frac{\tau u_x}{2}$ across a motionless disk of radius $r_{\text{asp}}$. The temperature of the half-spaces infinitely far away from the contact area are given by $T_{\text{b}}$.

# S6 – Conductivities of PAO4 and DLC in the TCM

**Table S4.** Conductivities of the lubricant $k_{PAO4}$ and the substrate $k_{DLC}$ used in the TCM model in Figure 10.

| Systems in panels a-c | $k_{PAO4}$ (Wm$^{-1}$K$^{-1}$) | $k_{DLC}$ (Wm$^{-1}$K$^{-1}$) |
|---|---|---|
| $u_x$ = 10 m/s | 0.0814 | 1.2398 |
| $u_x$ = 50 m/s | 0.0573 | 3.5741 |
| $u_x$ = 100 m/s | 0.0576 | 4.6375 |
| System in panel d | $k_{PAO4}$ (Wm$^{-1}$K$^{-1}$) | $k_{DLC}$ (Wm$^{-1}$K$^{-1}$) |
| $u_x$ = 10 m/s, $P$ = 1 GPa | 0.0427 | 1.2242 |

The values of the conductivities used for panels a-c in Figure 10 were calculated for each sliding velocity as the average value for all pressures ($P$ = 0.2, 0.5, 1.0, 1.5 and 2.0 GPa) by fitting the corresponding MD (TW) data with Eqs. 12 and 14. For panel d, the values were taken directly from the fit of the TW system at $u_x$ = 10 m/s, $P$ = 1 GPa. The fitted values of $k_{PAO4}$ underestimate experimental values measured for 4 cSt bulk PAO[7], as thermal conductivity in nanoconfined PAO4 is strongly reduced[7]. The thermal conductivities $k_{DLC}$ in our MD simulations are in the same range as those measured in real DLC[8,9].

# S7 – Temperature profiles in the TW systems

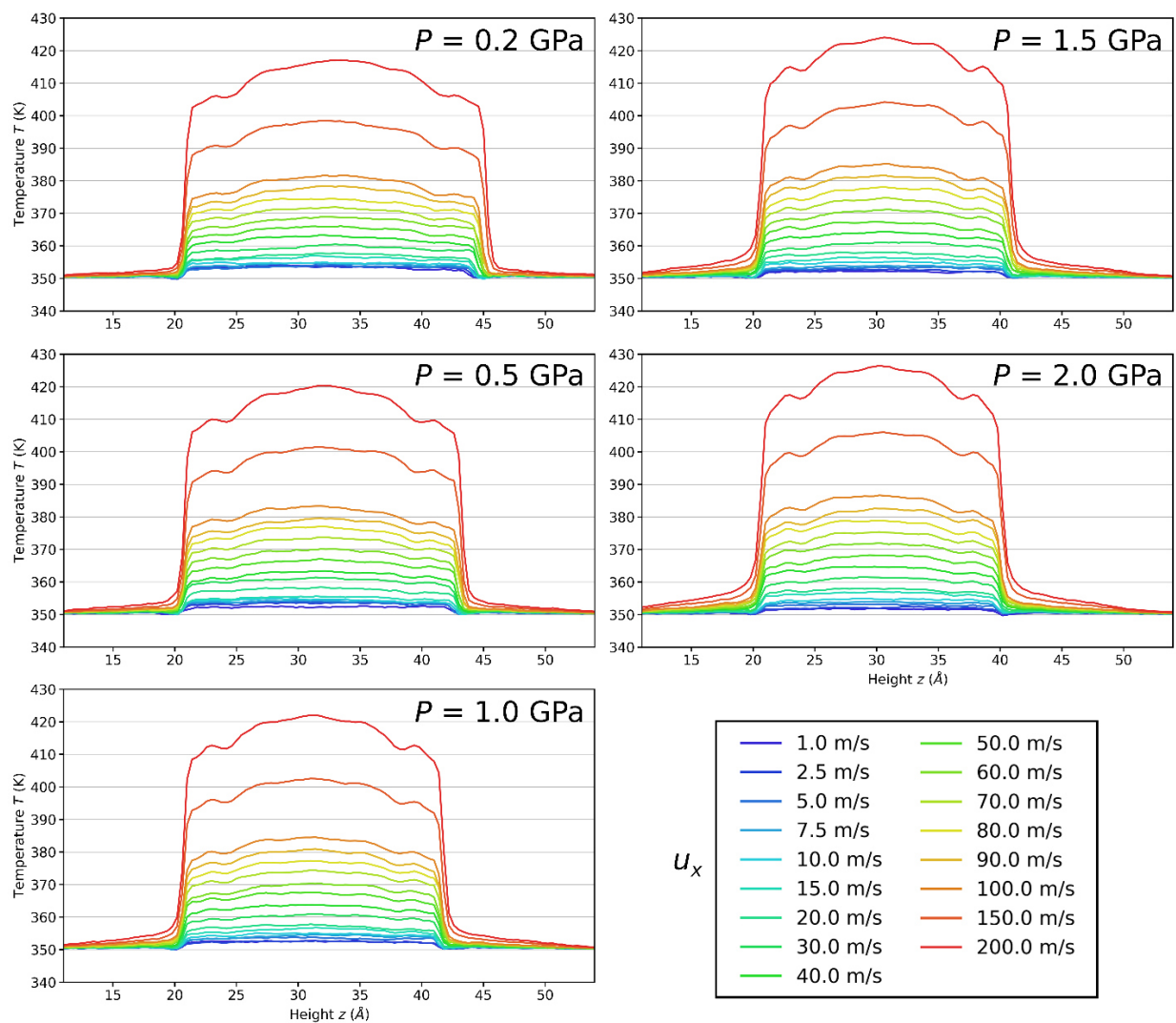

**Figure S10.** Temperature profiles for all the simulations with the TW scheme.

# S8 – Temperature of the lubricant and slip velocity as a function of the sliding velocity in the TW systems

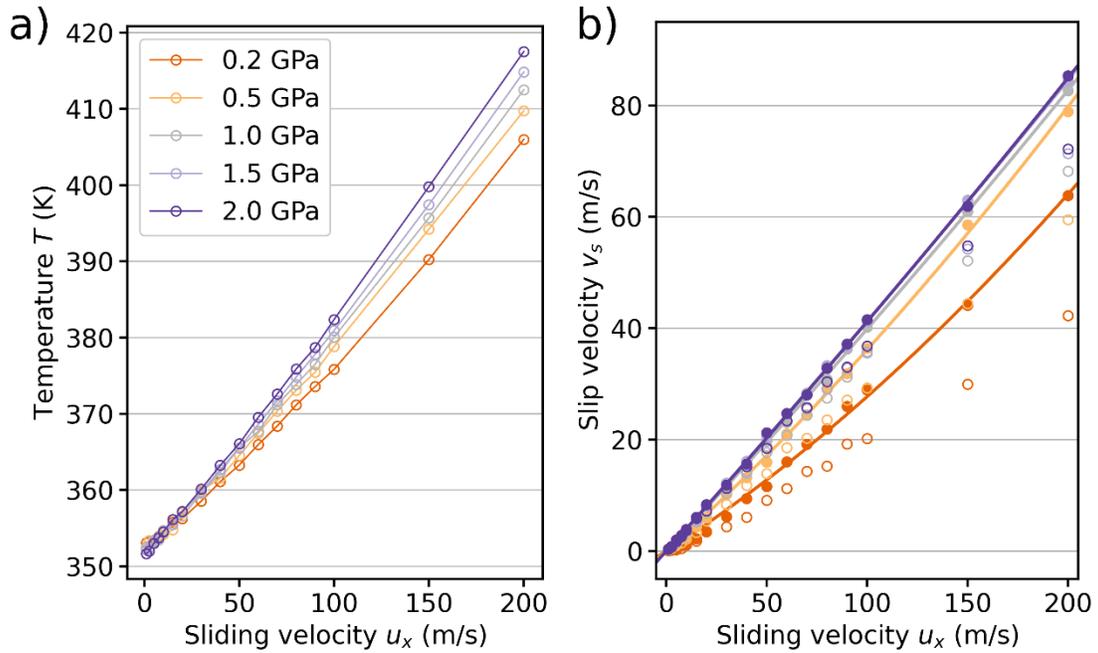

**Figure S11.** (a) Temperature of the fluid in contact with the wall in the TW system as a function of the sliding velocity (temperature is constant in the TWTF system). The line connecting the data points is a guide to the eye. (b) Slip velocity in the TW system (empty circles) compared to the TWTF system (filled circles) as a function of the sliding velocity. The line is a parabolic fit of the TWTF data points.